\PassOptionsToPackage{prologue,dvipsnames}{xcolor}
\documentclass[acmsmall,screen]{acmart}

\usepackage{listings}
\usepackage{tcolorbox}
\usepackage{cleveref}
\usepackage{url}
\usepackage{multirow}
\usepackage[dvipsnames]{xcolor}
\usepackage{adjustbox}  %
\usepackage{tabularx}   %
\usepackage{colortbl}
\usepackage{makecell} %
\usepackage{graphicx} %
\usepackage{wrapfig}

\newcolumntype{W}[1]{>{\centering\arraybackslash}wc{#1}}
\newcolumntype{G}[1]{>{\columncolor[HTML]{E8E8E8}\centering\arraybackslash}wc{#1}}

\usepackage[normalem]{ulem} %
\usepackage{xspace} %
\definecolor{colgray}{gray}{0.75}

\usepackage{ifthen}
\newboolean{showcomments}
\setboolean{showcomments}{true} %
\ifthenelse{\boolean{showcomments}}
  {\newcommand{\nb}[2]{
    \fcolorbox{gray}{yellow}{\bfseries\sffamily\scriptsize#1}
    {$\blacktriangleright$#2$\blacktriangleleft$}
   }
   
  }
  {\newcommand{\nb}[2]{}
   
  }

\newcommand{\Tratto}{\textsc{Tratto}\xspace}
\newcommand{\tratto}{\textsc{Tratto}\xspace} %
\newcommand{\njdconditions}{3,150\xspace}

\newcommand{\nemptyoracles}{23,397\xspace}

\newcommand{\collector}{\emph{Token collector}\xspace}
\newcommand{\filter}{\emph{Token filter}\xspace}

\newcommand{\evaluator}{\emph{Oracle evaluator}\xspace}
\newcommand{\selector}{\emph{Token selector}\xspace}

\def\|#1|{\mathid{#1}}
\newcommand{\mathid}[1]{\ensuremath{\mathit{#1}}}
\def\<#1>{\codeid{#1}}
\protected\def\codeid#1{\ifmmode{\mbox{\smaller\ttfamily{#1}}}\else{\smaller\ttfamily #1}\fi}

\AtBeginDocument{%
  }

\setcopyright{acmlicensed}
\acmDOI{10.1145/3728960}
\acmYear{2025}
\acmJournal{PACMSE}
\acmVolume{2}
\acmNumber{ISSTA}
\acmArticle{ISSTA083}
\acmMonth{7}
\received{2024-10-31}
\received[accepted]{2025-03-31}

\definecolor{custompurple}{rgb}{0.64,0.08,0.64}
\definecolor{customblue}{rgb}{0.13,0.67,0.8}
\definecolor{customred}{rgb}{0.8,0.16,0.16}
\definecolor{customgrey}{rgb}{0.9,0.9,0.9}
\definecolor{pgrey}{rgb}{0.46,0.45,0.48}

\newtcbox{\stepColorBox}[1][boxrule=0.5mm]{
 box align=base,
 nobeforeafter,
 colframe=black,
 colback=white,
 size=small,
 left=0pt,
 right=0pt,
 fontupper=\bfseries,
 colupper=black,
 halign=center,valign=center,square,circular arc,
 #1}

\lstdefinelanguage{ml-model-input}{
    backgroundcolor=\color{customgrey},
    stringstyle=\color{blue},
    keywordstyle=\color{violet}\bfseries,
    commentstyle=\color{green},
    keywords=[1]{token,tokenInfo,tokenClass,oracleSoFar,tokenClassesSoFar,oracleType,packageName,className,javadocTag,methodSignature,methodJavadoc},
    morecomment=[l]{//},
    morecomment=[s]{/*}{*/},
    escapeinside={(*@}{@*)}
}

\lstdefinestyle{listingtop}{
  float=tp,
  floatplacement=tbp,
  belowcaptionskip=-0.5cm,
}

\makeatletter
\newcommand\realnumberstyle[1]{}
\newcommand{\highlightlines}[3]{%
    {\realnumberstyle{#3}}%
    \begingroup
    \lst@basicstyle
    \ifnum\value{lstnumber}>2 %
    \ifnum\value{lstnumber}<4
        \color{#1}%
        \rlap{\hspace*{\lst@numbersep}%
        \color@block{1.022\linewidth}{\ht\strutbox}{\dp\strutbox}%
        }%
    \fi
    \fi
    \ifnum\value{lstnumber}>6 %
    \ifnum\value{lstnumber}<8
        \color{#1}%
        \rlap{\hspace*{\lst@numbersep}%
        \color@block{1.022\linewidth}{\ht\strutbox}{\dp\strutbox}%
        }%
    \fi
    \fi
    \ifnum\value{lstnumber}>3 %
    \ifnum\value{lstnumber}<6
        \color{#2}%
        \rlap{\hspace*{\lst@numbersep}%
        \color@block{1.022\linewidth}{\ht\strutbox}{\dp\strutbox}%
        }%
    \fi
    \fi
    \ifnum\value{lstnumber}>7 %
    \ifnum\value{lstnumber}<9
        \color{#2}%
        \rlap{\hspace*{\lst@numbersep}%
        \color@block{1.022\linewidth}{\ht\strutbox}{\dp\strutbox}%
        }%
    \fi
    \fi
    \ifnum\value{lstnumber}>22 %
        \color{#2}%
        \rlap{\hspace*{8.1\lst@numbersep}%
        \color@block{1.022\linewidth}{\ht\strutbox}{\dp\strutbox}%
        }%
    \fi
    \endgroup
}
\makeatother

\begin{document}

\title{\Tratto: A Neuro-Symbolic Approach to Deriving Axiomatic Test Oracles}

\author{Davide Molinelli}
\affiliation{%
  \institution{Constructor Institute, Università della Svizzera italiana}
  \country{Switzerland}}
\email{dm@constructor.org}
\orcid{0000-0002-3995-8529}

\author{Alberto Martin-Lopez}
\affiliation{%
  \institution{Università della Svizzera italiana}
  \country{Switzerland}}
\email{alberto.martin@usi.ch}
\orcid{0000-0001-5501-9225}

\author{Elliott Zackrone}
\affiliation{%
  \institution{University of Washington}
  \country{USA}}
\email{ezackr@cs.washington.edu}
\orcid{0009-0008-3071-0480}

\author{Beyza Eken}
\affiliation{%
  \institution{Università della Svizzera italiana}
  \country{Switzerland}}
\email{beyza.eken@usi.ch}
\orcid{0000-0002-6824-2765}

\author{Michael D. Ernst}
\affiliation{%
  \institution{University of Washington}
  \country{USA}}
\email{mernst@cs.washington.edu}
\orcid{0000-0001-9379-277X}

\author{Mauro Pezzè}
\affiliation{%
  \institution{Constructor Institute, Università della Svizzera italiana}
  \country{Switzerland}}
\email{mp@constructor.org}
\orcid{0000-0001-5193-7379}

\renewcommand{\shortauthors}{Molinelli et al.}

\begin{abstract}
  
  This paper presents \Tratto, a neuro-symbolic approach that generates assertions (boolean expressions) that can serve as axiomatic oracles, from source code and documentation.  
  The symbolic module of \Tratto takes advantage of the grammar of the programming language, the unit under test, and the context of the unit (its class and available APIs) to restrict the search space of the tokens that can be successfully used to generate valid oracles.  
  The neural module of \Tratto uses transformers fine-tuned for both deciding whether to output an oracle or
  not and selecting the next lexical token to incrementally build the oracle from the set of tokens returned by the symbolic module.
  Our experiments show that \Tratto outperforms the
  state-of-the-art axiomatic oracle generation approaches, with 73\% accuracy, 72\% precision, and 61\% F1-score, largely higher than the best results of the symbolic and neural approaches considered in our study (61\%, 62\%, and 37\%, respectively). \tratto can generate three times more axiomatic oracles than current symbolic approaches, while generating 10 times less false positives than GPT4 complemented with few-shot learning and Chain-of-Thought prompting.
\end{abstract}

\ccsdesc[500]{Software and its engineering~Software testing and debugging}
\ccsdesc[500]{Computing methodologies~Neural networks}
\ccsdesc[300]{Computing methodologies~Natural language processing}

\keywords{test oracle, automated oracle generation, deep learning, transfer learning}

\maketitle

\section{Introduction}
\label{sec:intro}

Software testing is an expensive activity.
While many testing activities can be automated, for example, test
execution with JUnit~\cite{cheon:jmlJunit:ECOOP:2002}, and unit test input
generation with Randoop~\cite{Pacheco:Randoop:ICSE:2007} and
EvoSuite~\cite{Fraser:Evosuite:TSE:2013}, other activities still require significant human effort. 
The oracle problem is still largely open and its automation is limited~\cite{barr:survey:tse:2015,Pezze:Oracles:Advances:2015}.
Current approaches generate implicit and regression oracles, which address only some aspects of the oracle problem: 
implicit oracles detect crashes and exceptions, while regression oracles detect differences in behavior between different versions.  

In this paper, we focus on the automatic generation of \emph{axiomatic
oracles}. The current approaches to generate test oracles produce
\textit{concrete oracles}, that is, assertions on specific concrete
inputs. For instance, \codeid{assertTrue(sum(20,22)==42)}
indicates the expected result for the concrete input values 20 and 22.  
Axiomatic oracles generalize concrete oracles as assertions on the inputs and outputs variables and state. 
For instance \codeid{assertTrue(sum(a,b)==a+b)} predicates on the expected result for all pairs of inputs.
Axiomatic oracles are essential elements in tools such as QuickCheck~\cite{ClaessenH2000} and jmlunit~\cite{CheonL2002} as an aid for testing. Axiomatic oracles are core assets of
property-based testing~\cite{FinkB1997}, metamorphic relations~\cite{ChenCY1998},
parameterized unit tests~\cite{TillmanS2005}, and
theories~\cite{Saff2007:OOPSLA2007demo}.

The benefits of axiomatic oracles are particularly significant when dealing
with large, automatically generated test suites. Tools like
Randoop~\cite{Pacheco:Randoop:ICSE:2007} and
EvoSuite~\cite{Fraser:EvoSuite:ESECFSE:2011} generate test cases
that rely on implicit and regression oracles only, and may result in false alarms. Generating concrete oracles manually for each test input to filter out false alarms is extremely effort-demanding and impractical for large test suites.
Axiomatic oracles in the form of preconditions, normal and exceptional postconditions within the unit under test are independent from the number of test cases, and can effectively reduce false alarms.   
Preconditions rule out invalid program inputs, thus reducing false positives. Normal and exceptional postconditions identify failures, thus reducing false negatives.

  Beyond testing, axiomatic test oracles are useful for other software engineering (SE) tasks. For example, they aid program comprehension~\cite{kanstren2009program,cornelissen2009systematic} by specifying expected behavior unequivocally, unlike ambiguous natural language descriptions. They support requirements engineering~\cite{cheng2007research} by providing formal specifications to validate and ensure completeness and consistency. In program synthesis~\cite{parisotto2016neuro}, axiomatic oracles can guide the generation of correct code by offering clear specifications. Lastly, in runtime verification~\cite{leucker2009brief,wang2018oracles}, they serve as properties to monitor, ensuring the program meets its specifications during execution.

Current approaches for generating test oracles are either symbolic- or
neural-based.
Symbolic approaches derive axiomatic oracles, such as program
invariants~\cite{Ernst:Daikon:SCP:2007}, metamorphic
relations~\cite{Blasi:memo:JSS:2021}, and
postconditions~\cite{Blasi:Jdoctor:ISSTA:2018} from formal
specifications~\cite{antoy:testOracles:TSE:2000,day:oraclesSpecifications:SOFTAIR:1985}, software documentation~\cite{Peters:OraclesFromDocumentation:TSE:1998}, and program
traces~\cite{Wang:Oracles:Forte:2004} using a pre-defined set of rules and patterns.
Neural approaches leverage deep learning (DL)~\cite{Dinella:TOGA:ICSE:2022} and transfer learning~\cite{mastropaolo:learning:press:2023}
to generate either concrete test cases~\cite{Tufano:TestCase:2021} or
concrete oracles for a given test
input~\cite{Dinella:TOGA:ICSE:2022,Watson:AssertionGenerationNN:ICSE:2020}.
Symbolic and neural approaches both suffer from limitations. Neural
approaches are well suited to handle fuzziness, for example, when deriving
test oracles from ambiguous natural language descriptions.  However, they 
require large amounts of (usually labeled) data and they are expensive to
run. Symbolic techniques perform well by leveraging a fixed set of
domain-specific rules, for instance, pattern and lexical
matching~\cite{Blasi:Jdoctor:ISSTA:2018,Blasi:memo:JSS:2021,Blasi:camema:ASE:2022},
but do not generalize beyond the hard-coded rules.
To the best of our knowledge, the potential of neural approaches for generating widely applicable axiomatic test oracles is still unexplored.

This paper introduces \Tratto, a neuro-symbolic approach to derive
axiomatic oracles from commonly available software artifacts such as source
code and documentation. An axiomatic oracle is composed of lexical
tokens.\footnote{Unless otherwise specified, we use ``token'' in the
traditional sense of compilers,
lexers, and programming languages.  Machine learning
transformers use ``token'' for a different concept (e.g., the \emph{lexical} token \texttt{someVar} may be \emph{tokenized} into two \emph{transformer} tokens, \texttt{some} and \texttt{Var}).}
For instance, the axiomatic oracle ``\codeid{result>0;}'' is
composed of four tokens, `\codeid{result}',  `\codeid{>}', `\codeid{0}', and `\codeid{;}'. 
\Tratto (\underline{TRA}nsformer-based
\underline{T}oken-by-\underline{T}oken \underline{O}racle generation)
reformulates the oracle generation problem as a token generation
problem.

\Tratto generates oracles token by token.
At each token-generation iteration, \Tratto restricts the search
space of the next possible tokens in two ways.  The first restriction is
based on a programming language grammar and the portion of the oracle
generated so far.  For instance, a boolean expression cannot be the argument with a
``\codeid{>}'' operator.
The second restriction is based on the symbols that are in scope.
These symbols include the method parameters and the
return value (for postconditions), fields and methods both in the current class and accessible through them.

\Tratto implements a neural approach  to select a token from the set of available tokens.
\Tratto leverages pre-trained transformers fine-tuned on the task of selecting candidate tokens based on (i)~the oracle generated so far, (ii)~the unit under test (including source code and documentation), and (iii)~additional unit context, for instance, information about the available APIs.

The neural module of \tratto features a multitask model trained on a dataset that was expressed in two different ways (for two different tasks): a dataset of oracles, and a dataset of tokens that appear in the oracles.  The dataset substantially extends the dataset available in the replication package of Blasi et al.~\cite{Blasi:Jdoctor:ISSTA:2018,jdoctor-dataset}.
We enhanced the initial dataset by (i)~fixing semantically-incorrect
oracles, (ii)~adding oracles that we could infer from the code and the
documentation,
(iii)~adding oracles from other publicly available Java projects, and
(iv)~automatically augmenting the Javadoc comments with semantically
equivalent Javadoc comments that refer to the same oracles. 
The resulting dataset features 34,249 oracle samples.
We produced a dataset of 188,900 tokens from the tokens in the oracles.
Both datasets are publicly available in multiple formats, suitable for training
sequence-to-sequence and classification models, to support future
research~\cite{supplementarymaterial}.

We experimentally evaluated \Tratto.
We compared the performance of
different code models to support the neural module, Code Gemma
(Google)~\cite{codegemmateam2024codegemmaopencodemodels}, StarCoder2
(BigCode)~\cite{lozhkov2024starcoder2stackv2} and Code Llama
(Meta)~\cite{rozière2024codellamaopenfoundation}, all in their version
featuring 7B trainable parameters. We selected Code Llama since it
outperforms the other models for this task, with 91\% accuracy in
predicting the next oracle token.

We performed
two ablation studies to compare \tratto with (i)~a purely neural model
trained for predicting oracles, not supported by the token-by-token
symbolic approach, and (ii)~a version of \tratto featuring two different
models, instead of a single multitask model. The results show that the
symbolic module and the multitask model provide an additional +6 and +3
percentage points of accuracy in predicting oracles, respectively.

We compared \Tratto with two state-of-the-art symbolic and neural approaches for axiomatic oracle generation, Jdoctor (symbolic)~\cite{Blasi:Jdoctor:ISSTA:2018} and GPT4 (neural)~\cite{OpenAI:GPT4-o:corr:2024} on a ground-truth dataset of axiomatic test oracles. 
The experimental results that we discuss in Section~\ref{sec:results} indicate that \tratto outperforms both Jdoctor and GPT4 in terms of \emph{accuracy} (73\% for \tratto, 61\% for Jdoctor, 40\% for GPT4), \emph{precision} (72\% for \tratto, 62\% for Jdoctor,  24\% for GPT4), and \emph{F1-score} (61\% for \tratto, 25\% for Jdoctor, 37\% for GPT4).  \tratto's \emph{recall} (52\%) is better than Jdoctor's (16\%) and worse than GPT4's (89\%).
In overall terms, \tratto achieves an excellent balance between generating oracles ($3\times$ more than Jdoctor) while incurring in few false positives ($10\times$ less than GPT4).

  We also compared the robustness of \tratto, Jdoctor and GPT4 in generating correct oracles when the documentation of the units under test was modified in different ways (details in Section~\ref{sec:rq4}). The results show that \tratto and GPT4 have comparable performance, generating 195 and 208 correct and compilable oracles, respectively, from a set of 220 descriptions derived from 55 original descriptions. Jdoctor generated only 49 correct and compilable oracles.

  Lastly, we evaluated the effectiveness of the oracles generated by \tratto and Jdoctor for increasing the mutation score of test suites automatically generated with EvoSuite~\cite{Fraser:EvoSuite:ESECFSE:2011}. We could not evaluate GPT4 for this purpose due to the large amount of non-compilable oracles generated (552 out of 1,213), which made it impossible to automate the insertion of oracles into the test cases. We considered two types of test suites: with implicit oracles only, and with both implicit and regression oracles.   
  From among 6 projects considered, \tratto increased the mutation score of 5 test suites with implicit oracles and 3 test suites with both implicit and regression oracles, while Jdoctor increased the mutation score of 4 test suites in the former case and in no cases in the latter.

In summary, this paper discusses the limitations of the state-of-the-art
approaches to generate axiomatic oracles (Section~\ref{sec:example}), and
makes the following contributions:
\begin{enumerate}
    \item It defines a novel approach that iteratively generates test oracles, token by token, by combining a symbolic with a neural approach to steer the generation of tokens toward valid oracles, thus reducing the impact of false positives of purely neural approaches (Section~\ref{sec:tratto-architecture}).  
    \item It introduces \Tratto, a neuro-symbolic approach and corresponding tool to generate axiomatic oracles from source code and documentation (Sections~\ref{sec:symbolic-module} and \ref{sec:neural-module}).
    \item It proposes a collection of comprehensive datasets of oracles and tokens that can be reused for training future models for generating oracles (Section~\ref{sec:data-collection}). 
    \item It performs empirical studies to: (i)~compare the performance of different code models for the tasks of oracle evaluation and token selection; (ii)~highlight the contributions of \tratto's components to its overall performance; and (iii)~compare \tratto with state-of-the-art approaches for oracle generation, showing its superior performance in terms of correctness, robustness and applicability to enhancing automatically generated test suites (Section~\ref{sec:evaluation}).
\end{enumerate}

\section{Motivating Example}
\label{sec:example}

\lstset{
  frame=bt,
  numbers=left,
  xleftmargin=12pt,
  xrightmargin=5pt,
  frame=single,
  framesep=0pt,
  framexleftmargin=3pt,
  framexrightmargin=3pt,
  commentstyle=\color{ForestGreen}
}

\begin{wrapfigure}{r}{0.5\textwidth}
\vspace{-0.5cm}
\begin{lstlisting}[language=java, caption={Documentation and implementation of 
      method \href{https://github.com/jfree/jfreechart/blob/e2d6788d594c51941ddae337ae666fda5c52ad9f/src/main/java/org/jfree/chart/renderer/category/AbstractCategoryItemRenderer.java\#L256}{setSeriesItemLabelGenerator} from \href{https://github.com/jfree/jfreechart/tree/e2d6788d594c51941ddae337ae666fda5c52ad9f}{JFreeChart}.},
    label=list:precondition, numberstyle=\tiny\color{pgrey}]
/**
 * Sets the item label generator for a series and
 * sends a {@link RendererChangeEvent} to all
 * registered listeners.
 *
 * @param series  the series index (zero based).
 * @param generator  the generator
 * (<code>null</code> permitted).
 *
 * @see #getSeriesItemLabelGenerator(int)
 */
public void setSeriesItemLabelGenerator(int series,
        CategoryItemLabelGenerator generator) {
    setSeriesItemLabelGenerator(series, generator,
            true);
}
\end{lstlisting}
\vspace{-0.65cm}
\end{wrapfigure}

This section shows how state-of-the-art oracle generators perform on an example method (Listing~\ref{list:precondition}). Their limitations motivate \tratto.

The documentation in Listing~\ref{list:precondition} describes a precondition
``\codeid{series >= 0}'' in natural language (\emph{``the series index (zero based)''}, line 6).
If a test generator produces a unit test case with a negative
\codeid{series} as argument, the test crashes.  The precondition indicates
that the test case is invalid, and avoids executing the test with a false positive result.

The main state-of-the-art oracle generators do not generate this precondition: 

\noindent
\textbf{TOGA}~\cite{Dinella:TOGA:ICSE:2022}, a neural
approach for generating test assertions, does not generate
preconditions at all.

\noindent
\textbf{Jdoctor}~\cite{Blasi:Jdoctor:ISSTA:2018}, a symbolic approach, does not generate the precondition either, since it is based on a set of patterns and rules that do not match the comment.  

\noindent
\textbf{EvoSuite}~\cite{Fraser:EvoSuite:ESECFSE:2011}, a search-based test generator, generates regression test oracles, and no preconditions.

\noindent
\textbf{GPT4}~\cite{gpt4}, a large language model (LLM) trained on natural language and code, correctly generates the precondition, but also generates wrong, useless, or non-compilable oracles such as ``\codeid{generator == null || generator != null}'' and ``\codeid{generator == null || generator is a valid CategoryItemLabelGenerator instance}''.

\noindent
\textbf{\Tratto} derives the precondition ``\codeid{series >= 0}'' that prevents the generation of invalid test cases and false positives, and does not generate wrong oracles like GPT4. The neural module of \tratto generalizes to previously unseen oracles, beyond  the fixed set of rules and patterns of Jdoctor, while the symbolic module restricts the possible oracles to compilable oracles.

\section{\Tratto}
\label{sec:tratto}

\Tratto generates  
executable preconditions, regular and exceptional postconditions that together comprise axiomatic oracles.
\Tratto generates executable assertions in Java from Java source code and Javadoc comments.
Preconditions, like \emph{``param1 cannot be null''}, constrain the validity of the test input.
Postconditions, like \emph{``the output must be positive''}, and exceptional postconditions, like \emph{``if param1 is null, an exception is thrown''}, constrain the expected behavior of a program.

The input of \Tratto is the source code of the method under test and its
context, such as fields and methods in its class.
\tratto tries to generate a precondition from each
 \codeid{@param} tag,
 a regular postcondition from each \codeid{@return} tag, an exceptional
 postcondition from each \codeid{@throws} and \codeid{@exception} tag, and preconditions, regular and exceptional postconditions from the free-text part of the Javadoc, the method signature and the implementation.

\subsection{Architecture}
\label{sec:tratto-architecture}

\label{def:collector}
\label{def:filter}

\tratto integrates a symbolic and a neural module to generate test oracles iteratively, token by token. The symbolic module features a \collector and a \filter to restrict the possible tokens forming the oracle. The neural module consists of a multitask model that works in two modes: \evaluator and \selector. The former decides whether to generate an oracle or not, while the latter selects the tokens that incrementally form an oracle, based on those returned by the symbolic module.

\begin{wrapfigure}{r}{0.67\textwidth}
  \centering
  \includegraphics[width=0.67\textwidth]{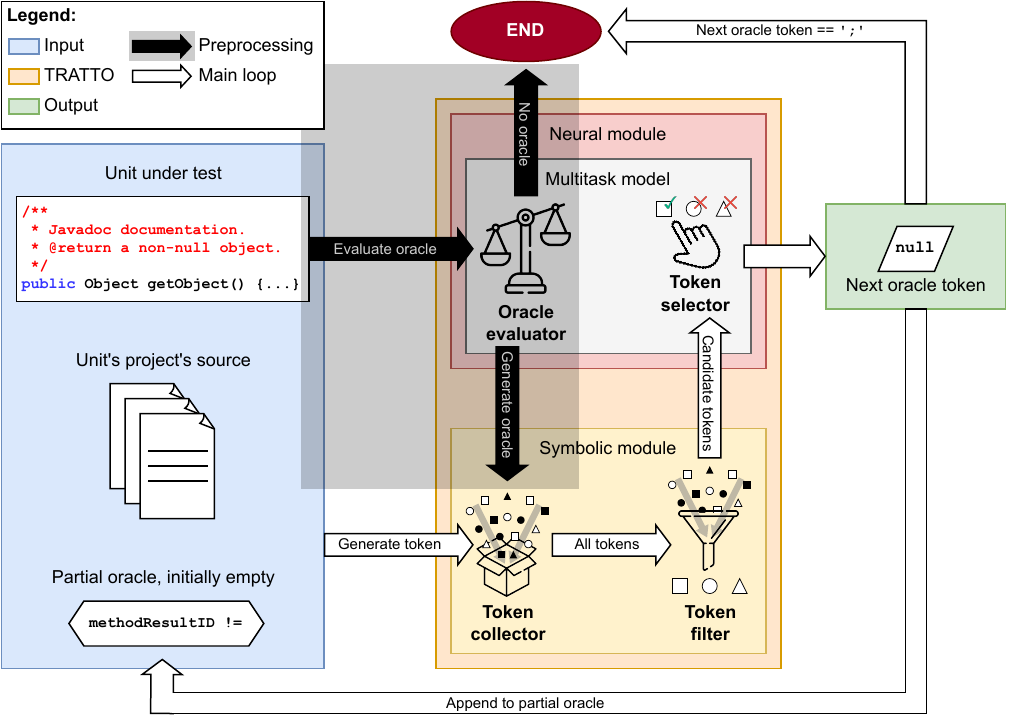}
  \caption{Workflow of \Tratto.
  }
  \label{fig:tratto}
\end{wrapfigure}

Figure~\ref{fig:tratto} shows the workflow of \Tratto. 
As a preprocessing step (black arrows), the \evaluator component of the neural module decides if an oracle should be generated, solely based on the documentation and code of the \emph{unit under test}. If so, it triggers the token-generation workflow (white arrows).
At each iteration, \tratto generates the next token of the oracle based on (i)~the unit under test,
(ii)~the \emph{unit project's source} (documentation and source of the whole project), and (iii)~the \emph{partial oracle} (portion of the oracle built so far).
\Tratto starts with an empty partial oracle and terminates when the next generated token
is a semicolon (`\codeid{;}').

The symbolic and the neural modules of \Tratto cooperate to incrementally generate the tokens that comprise the partial oracle, until they produce a complete oracle.
At each iteration, the \collector retrieves all possible tokens to build an oracle, for instance, parameters of the unit under test with
their fields and methods. Then, the \filter selects the subset of tokens
that are \emph{valid}, i.e., that can be added to the partial oracle to eventually
generate a
\emph{well-formed}
assertion (\emph{Candidate tokens} in Figure~\ref{fig:tratto}). For example, the \codeid{instanceof} operator may not follow a variable of primitive type, although it is a syntactically valid expression. Given a set of candidate tokens, the \selector of \tratto's neural module selects the next token that shall be added to the partial oracle.

\Tratto uses a novel token-level approach to generate oracles.
The state-of-the-art approaches like TOGA~\cite{Dinella:TOGA:ICSE:2022} and
Jdoctor~\cite{Blasi:Jdoctor:ISSTA:2018} generate a whole oracle at once.
Generating test oracles token-by-token brings two main benefits:

\textbf{(1)}
    The grammar-based symbolic approach of \Tratto: (i) generates only
     oracles that are compilable by
    construction, while neural-based approaches may generate oracles that do not compile~\cite{Tufano:TestCase:2021,gpt4}; (ii) generates any oracle that is consistent with the grammar, while some neural approaches generate only oracles compatible with a set of candidate assertions~\cite{Dinella:TOGA:ICSE:2022}; and (iii) collects contextual information from the unit under test useful to generate an oracle, for instance, possible method calls to external libraries, calls that neither symbolic nor neural approaches can access without manual intervention, while prompt-based neural approaches require manually specified data.
    We define the grammar-based symbolic approach of \Tratto in Section~\ref{sec:symbolic-module}.

\textbf{(2)}
The neural model: (i) is trained  on a dataset of tokens much larger than a dataset of oracles, since oracles are composed of multiple tokens; and (ii) takes full advantage of transformers capabilities of learning patterns and associations related to natural language and source code, by relying on training with significantly more data, beyond the limitations of semantic approaches that rely on pattern, lexical and semantic matching.
    We discuss the  neural module of \tratto in
    Section~\ref{sec:neural-module} and the benefits of the tokens dataset
    in Section~\ref{sec:data-collection}.

\subsection{Symbolic Module}
\label{sec:symbolic-module}

The symbolic module of \Tratto restricts the search space of the possible oracles for a given unit, to a set of tokens that produce compilable oracles, by construction. The \collector collects all possible tokens that could be used to form an oracle. The \filter discards the illegal tokens (syntactically and semantically invalid tokens that would make the oracle non-compilable) with a grammar-based approach.

\subsubsection{Token Collector}
\label{sec:token-collector}
The \collector creates a set of all the tokens that \Tratto needs to generate
oracles for the target unit. These tokens may be \emph{generic} tokens related to the target unit under test, collected in the first token-generation iteration, or \emph{specific} tokens related to the current partial oracle, collected only in some iterations.

In the first iteration (when the partial oracle is empty), the \collector gathers three types of \emph{generic} tokens: (i) common tokens which could be
part of any oracle, such as operators, keywords, and common constants like
0 and 1; (ii) tokens extracted from the project under test, including classes and their
respective fields and methods (e.g., \codeid{CollectionUtils.isEmpty()}); and (iii) tokens
extracted from the method under test, including its parameters (if any) as well as fields and methods callable upon those parameters, containing class, and return value of the method (e.g., \codeid{this.\textbf{contains}(o)}).

At each subsequent iteration, the \collector may augment the set of tokens with \emph{specific} tokens that may occur when the last token of the partial oracle is a period that
follows an expression that is a class of the project,
\codeid{this}, \codeid{methodResultID} (which identifies the return value
of the method under test), or a method parameter.
For instance the \collector adds the tokens `\codeid{hasNext}' and `\codeid{next}' to the set of tokens when the partial oracle ends with ``\codeid{this.iterator().}''.

\subsubsection{Token Filter}
\label{sec:token-filter}
The \filter discards both the tokens that are syntactically illegal in the next position of the partial oracle (for instance, `\codeid{true}' is not syntactically correct as the next token after ``\mbox{\codeid{arg1 >}}''), and the tokens that would result in a compilation error of the oracle (for instance, the \codeid{>} operator may follow only an expression that evaluates to a numeric type). 
The \filter prunes the set of tokens with a grammar-based approach.
\Tratto also implements several \emph{context restrictions} that discard
tokens that may not occur in the next position of the oracle, even if
grammatically legal, for instance, \codeid{methodResultID} may not be used
if the method under test is void.

\vspace{0.2cm}\noindent\textbf{\Tratto Grammar.}
Axiomatic oracles are boolean expressions. \tratto implements a custom-defined grammar to express oracles (available in the replication package~\cite{supplementarymaterial} due to space restrictions).
In a nutshell, the grammar supports
expressions consisting of variables potentially chained with class fields
and method calls, such as ``\codeid{this.a.b()}'';
arithmetic operations, as in ``\<arg2+arg3>'';
comparison operators, as in ``\codeid{methodResultID != null}'';
and conjunctions and disjunctions formed with the 
logical operators \codeid{\&\&} and \codeid{||}.
This simple but effective grammar can represent all procedure specifications from the work by Blasi et al.~\cite{Blasi:Jdoctor:ISSTA:2018}, while permitting new oracles.

\vspace{0.2cm}\noindent\textbf{Context Restrictions.}
A grammar does not suffice to generate a test oracle, since it lacks the
underlying semantics of the specific tokens conforming it. For example,
according to the \Tratto grammar, the token `\codeid{instanceof}' is valid
after the partial oracle ``\codeid{methodResultID}''.  However, if the type of the
variable \codeid{methodResultID} (i.e., the return type of the method under
test) is primitive, the resulting expression would not compile. We refer to cases like this as \emph{context restrictions}, since they restrict tokens depending
on the context of the oracle (so far). \Tratto implements 27 context restrictions (documented in the replication package~\cite{supplementarymaterial}).
The context restrictions also alleviate the load of the neural module, by reducing  the number of tokens that it must evaluate to generate  accurate oracles.

\subsection{Neural Module}
\label{sec:neural-module}
The neural module comprises the \evaluator---decides whether an oracle should be generated or not---and the \selector---guides the generation of oracles toward
optimal solutions, token by token, in all iterations. The neural module
features a multitask DL model trained for both tasks.

\subsubsection{Oracle Evaluator}
\label{sec:oracle-evaluator}
Although the oracle evaluation task is a binary task that determines whether an oracle should be generated or not, we treat it as a generation task, so that we can reuse the same auto-regressive model (e.g., specialized for code generation and completion) for both tasks. 
We ask the model to fill out the mask of a masked input. Listing~\ref{list:oracle-template} shows the input template used for this task.
It is composed of an oracle type and a Javadoc tag, if available (line~1), two next possible candidate tokens for the partial oracle (line~2), the mask token to fill out (line~4), and the method under test (lines~6--8). 
The model selects one of the two next possible tokens, either `\codeid{assertTrue(}' (an oracle should be generated) or `\codeid{//~No~assertion~possible}' (no oracle).

\subsubsection{Token Selector}
\label{sec:token-selector}
We treat the token selection task as a generation task, as for the \evaluator. Listing~\ref{list:token-template} shows the input template that we use for this task. With respect to the input of the \evaluator, this is augmented with (i) a list of next possible tokens and not just a binary choice (line~2), (ii) the partial oracle that is not empty (line~4), and (iii) the additional context with information about method signatures and field declarations corresponding to the next possible tokens (lines~10-11), aiming to provide more context to the model. 
The model \emph{selects} one of the next possible tokens that produces a new correct partial oracle when added to the tail of the input partial oracle.

\noindent\begin{minipage}[b]{0.5\textwidth}

\begin{lstlisting}[language=java, caption={Input template for the Oracle Evaluator.}, label=list:oracle-template, numberstyle=\tiny\color{pgrey}, mathescape=true]
// $\textbf{\textcolor{red}{<oracle\_type>}}$: "$\textbf{\textcolor{red}{<javadoc\_tag>}}$"
// Next possible tokens: ['assertTrue(', $\hspace{4em}$ '// No assertion possible']
// Assertion:
$\textbf{<FILL\_ME>}$

// Method under test:
$\textbf{\textcolor{red}{<method\_javadoc>}}$
$\textbf{\textcolor{red}{<method\_source>}}$
\end{lstlisting}

\end{minipage}
\hfill
\begin{minipage}[b]{0.5\textwidth} 

\begin{lstlisting}[language=java, caption={Input template for the Token Selector.}, label=list:token-template, numberstyle=\tiny\color{pgrey}, mathescape=true]
// $\textbf{\textcolor{red}{<oracle\_type>}}$: "$\textbf{\textcolor{red}{<javadoc\_tag>}}$"
// Next possible tokens: [$\textbf{\textcolor{red}{<next\_possible\_tokens>}}$]
// Assertion:
assertTrue($\textbf{\textcolor{red}{<partial\_oracle>}}$$\textbf{<FILL\_ME>}$

// Method under test:
$\textbf{\textcolor{red}{<method\_javadoc>}}$
$\textbf{\textcolor{red}{<method\_source>}}$

// Additional context:
$\textbf{\textcolor{red}{<method\_signatures\_and\_field\_declarations>}}$
\end{lstlisting}

\end{minipage}

Listing~\ref{list:token-input} is a complete example of the input to the model when acting as \selector. 
The example is a snapshot of the construction of an exceptional postcondition as shown in line~1 (partial oracle = ``\codeid{array.getClass().}'' in line~4). As the \codeid{getClass()} method returns an object of type \codeid{Class}, the next possible tokens are non-private fields and methods of class \codeid{Class} (line~2). The model selects  the correct token, in this case `\codeid{isArray}' (additional information provided in line 24), among the possible tokens, to replace the mask token in the partial oracle, thus continuing the oracle generation process.

\begin{wrapfigure}{r}{0.5\textwidth}
\vspace{-1.5cm}
\begin{lstlisting}[language=java, caption={Input to the model acting as Token Selector.}, label=list:token-input, numberstyle=\tiny\color{pgrey}, mathescape=true]
// Exceptional postcondition: "@throws IllegalArgumentException if <code>array</code> is${\xspace}$not an array."
// Next possible tokens: ['equals', 'toString', $\hspace{1em}$ 'isArray', 'getClassData', 'getClassLoader', ...]
// Assertion:
assertTrue(array.getClass().$\textbf{<FILL\_ME>}$

// Method under test:
/**
  * Constructs an ArrayListIterator that will
  * iterate over the values in the specified array.
  *
  * @param array the array to iterate over
  * @throws IllegalArgumentException if
  * <code>array</code> is not an array
  * @throws NullPointerException if
  * <code>array</code> is <code>null</code>
  */
public ArrayListIterator(final Object array){
    super(array);
}

// Additional context:
public boolean equals(Object arg0)
public String toString()
public native boolean isArray()
Object getClassData()
public ClassLoader getClassLoader()
...
\end{lstlisting}
\vspace{-0.7cm}
\end{wrapfigure}

\subsection{Data Collection}
\label{sec:data-collection}

We trained \tratto's neural model used for the \evaluator and \selector with datasets of oracles and tokens, respectively.
Figure~\ref{fig:data-collection} shows how we generated the datasets. We
started with an existing dataset of procedure
specifications~\cite{Blasi:Jdoctor:ISSTA:2018}, added comments with and without a corresponding oracle, and manually inspected and cleaned the
dataset.
We augmented this dataset via automated techniques.
We disaggregated oracles into tokens via the \collector and \filter components of \Tratto, to produce the tokens dataset.
Oracles and tokens datasets are available in the replication package~\cite{supplementarymaterial}.

\subsubsection{Procedure Specifications Dataset}
\label{sec:jdoctor-dataset}
Blasi et al.~\cite{Blasi:Jdoctor:ISSTA:2018} provide a dataset of \emph{procedure specifications} (preconditions, regular and exceptional postconditions) that correspond to Javadoc tags.
The procedure specifications express the intended behavior of a program and
therefore they can be used as test oracles. This dataset contains
\njdconditions tuples in the form of $\langle(u, \|jt|), o\rangle$,
where \emph{u} is the \emph{unit} under test, \emph{jt} is the
\emph{Javadoc tag} in the unit documentation, and \emph{o} is an executable Boolean expression that corresponds to the Javadoc tag and may be used as an \emph{oracle}. Blasi et al.~\cite{Blasi:Jdoctor:ISSTA:2018} provide also \nemptyoracles Javadoc tags for which their approach cannot generate axiomatic oracles.

\subsubsection{Oracles Dataset}
\label{sec:oracles-dataset}

\begin{figure}[t]
  \centering
  \includegraphics[width=0.89\textwidth]{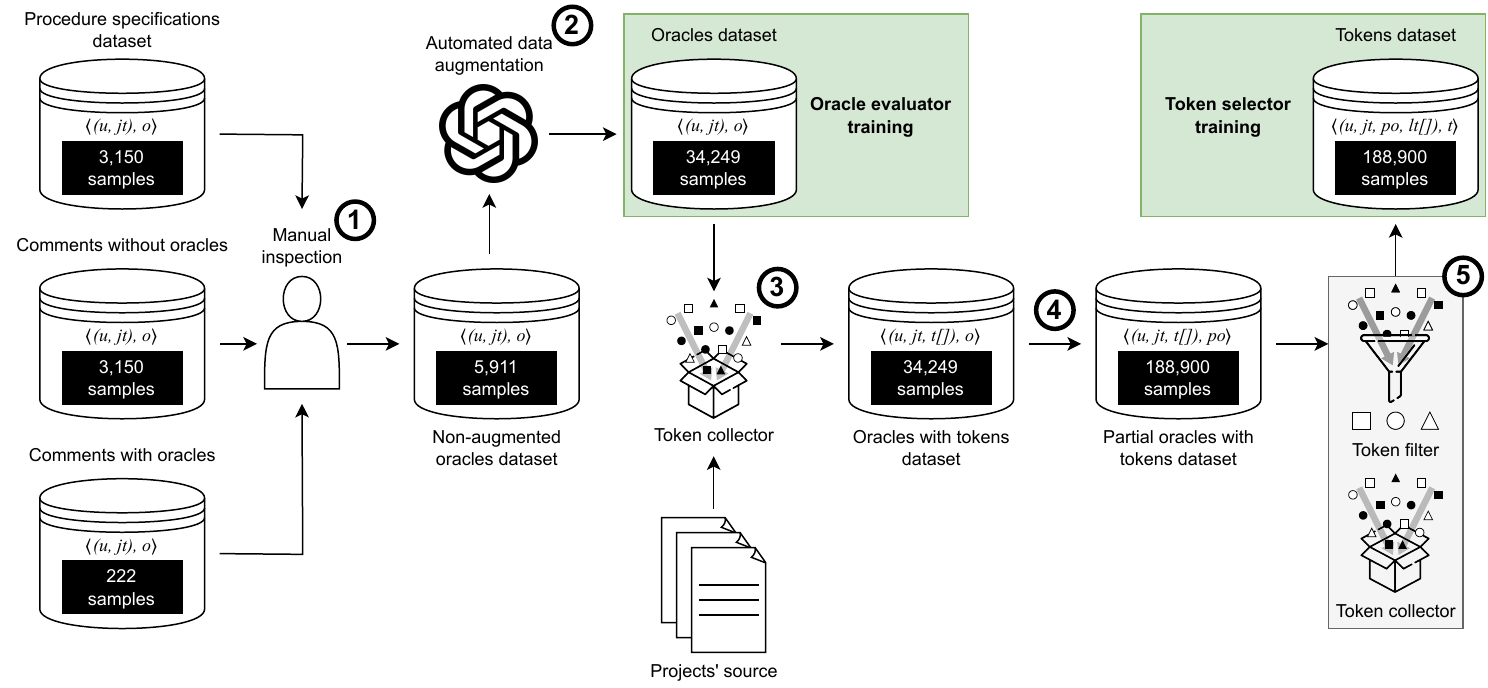}
  \caption{Data collection process.}
  \label{fig:data-collection}
  \vspace{-0.5cm}
\end{figure}

We created an initial oracles dataset by combining the \njdconditions procedure specifications from the dataset of Blasi et al.~\cite{Blasi:Jdoctor:ISSTA:2018} (\emph{positive instances}) with \njdconditions randomly sampled \emph{negative instances} (\emph{Comments without oracles} in Figure~\ref{fig:data-collection}) from among the \nemptyoracles Javadoc tags without oracles.
Since the authors report an average precision and recall of 92\% and 83\%, respectively, we decide to inspect the selected 6,300 instances (\njdconditions positive and negative) to fix possible mistakes (e.g., wrongly generated oracles).
Additionally,
we add 222 oracles from various Java projects on GitHub to increase the quality of the final dataset. This manual process (step \stepColorBox{1} in Figure~\ref{fig:data-collection}) leads to an initial oracles dataset of 5,911 samples, 4,582 of which are positive and 1,329 of which are negative. Indeed, a large number of the \njdconditions inspected comments without oracles resulted in oracles that could be derived, hence the larger number of positive instances in the resulting dataset.

\begin{wrapfigure}{r}{0.59\textwidth}
\vspace{-0.2cm}
\begin{lstlisting}[caption={Equivalent Javadoc tags generated with ChatGPT.}, label=list:javadoc-tags, numberstyle=\tiny\color{pgrey}]
@return the sum {@code a + b}. // Original
@return the total value of {@code a + b}
@return the result of adding {@code a} and {@code b}
@return the outcome of summing {@code a} and {@code b}.
@return the value obtained by adding {@code a} and {@code b}.
@return the sum of {@code a} and {@code b}
\end{lstlisting}
\vspace{-0.85cm}
\end{wrapfigure}

We augmented the oracles dataset by creating semantically equivalent
versions of the comments in the dataset with ChatGPT~\cite{openai2023chatgpt}.
We asked it to generate new Javadoc comments by suggesting equivalent versions of the comments, with a maximum of 5 (\stepColorBox{2}). 
Listing~\ref{list:javadoc-tags} shows a sample set of equivalent Javadoc tags generated with 
ChatGPT. Details of the prompts and data cleaning are available in the replication
package~\cite{supplementarymaterial}. 
We obtained 34,249 comments, 23,392 with an oracle (positive samples) and 10,857 without (negative samples).

\subsubsection{Tokens Dataset}
\label{sec:tokens-dataset}

Based on the oracles dataset, we generated an \emph{oracles with
tokens dataset}, by leveraging the
\collector to extract \emph{generic} tokens that could belong to each oracle (\stepColorBox{3}),
for instance, class names (e.g., \codeid{HashBiMap}), constants (e.g.,
\codeid{CollectionUtils.EMPTY\_COLLECTION}), and method names (e.g.,
\codeid{.toString()}). The \collector component (Section~\ref{sec:token-collector})
analyzes the source code of
the project related to the oracle to extract available tokens, and
produces a dataset of tuples in the form of $\langle$\emph{(u, jt, t[]),
o}$\rangle$, where \emph{t[]} refers to the \emph{list of all possible
tokens} that could be initially used to start building the oracle \emph{o}.

Subsequently, we automatically disaggregated oracles into partial oracles (\stepColorBox{4}) and used both the \collector and \filter to add relevant tokens and rule out invalid tokens for each partial oracle (\stepColorBox{5}), respectively, leading to a dataset of tokens in the form of $\langle (u, \|jt|, \|po|, \|lt|[]), t \rangle$ tuples, where
\emph{po} denotes a \emph{partial oracle}, \emph{lt[]} refers to the \emph{list of legal tokens} that could
possibly follow the partial oracle (e.g., `\codeid{0}', `\codeid{1}',
and `\codeid{SomeClass}' could follow ``\codeid{result >}''), and \emph{t} is the actual next token after
the partial oracle, which must be one of the tokens from the aforementioned list. The
tokens dataset contains 188,900 samples in total.

\begin{wrapfigure}{r}{0.5\textwidth}
  \vspace{-0.7cm}
  \centering
  \includegraphics[width=0.4\textwidth]{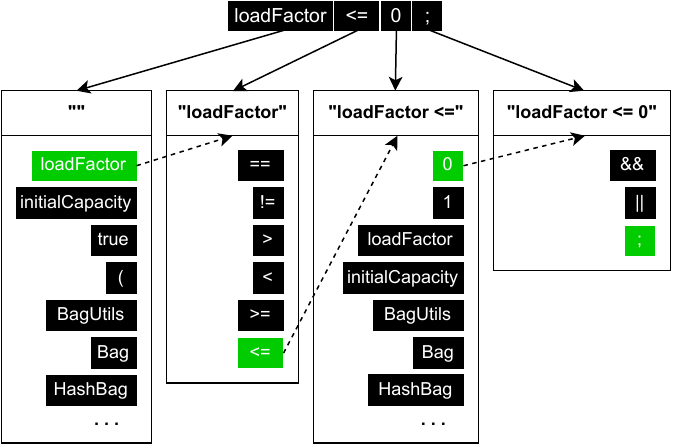}
  \caption{Converting one oracle sample into four token samples. Oracle ($o$) at the top; partial oracles ($po$) and legal tokens ($lt[]$) on top and bottom white boxes, respectively; next tokens ($t$) in green.}
  \label{fig:oracles2tokens}
  \vspace{-0.5cm}
\end{wrapfigure}

Figure~\ref{fig:oracles2tokens} shows an example of the conversion process
from an oracle sample to token samples. The oracle
``\codeid{loadFactor<=0;}''
represents an exceptional behavior, i.e., if the argument of the method
under test \codeid{loadFactor} is less than or equal to 0, an exception
should be thrown. This oracle contains four tokens, so the data generation
process produces four partial oracles.
For instance, in the first iteration, the oracle
may start with tokens such as parameters from the method under test (i.e.,
\codeid{loadFactor} and \codeid{initialCapacity}), an opening parenthesis,
the name of a class (e.g., if a static method is to be called), and so
on.
In the second iteration, when the partial oracle is
``\codeid{loadFactor}'', the next token can only be a comparison
operator
since \codeid{loadFactor} is of numeric type, thus the \filter discarded
all other tokens that may not be used in this context (including arithmetic operators which are not allowed in this particular position according to the \tratto grammar~\cite{supplementarymaterial}).
The same process is
repeated for the remaining tokens of the oracle.

\section{Evaluation}
\label{sec:evaluation}

Our experimental evaluation addresses the  following research questions:

\textbf{RQ\textsubscript{1}}: \emph{How do different LLM code models compare for evaluating oracles and selecting tokens?} We evaluate different code models aiming to find the most suitable one for the tasks performed by the neural module of \Tratto.

\textbf{RQ\textsubscript{2}}: \emph{What is the contribution of the symbolic module and the multitask model to the overall performance of \Tratto?} We carry out ablation studies to assess the performance of \Tratto without the symbolic module (a purely neural model not supported by the proposed token-by-token oracle generation approach) and without the multitask model (two different models for the \evaluator and \selector).

\textbf{RQ\textsubscript{3}}: \emph{What is the effectiveness of \Tratto in generating axiomatic test oracles and how does it compare with state-of-the-art neural and symbolic techniques?} We devise a ground-truth dataset of axiomatic test oracles to evaluate \Tratto, and compare it against state-of-the-art symbolic- and neural-based approaches for oracle generation.

\textbf{RQ\textsubscript{4}}: \emph{How robust is \tratto to documentation variations?} Based on the results of RQ\textsubscript{3}, we select the oracles that all approaches correctly predict, we apply systematic variations to their associated documentation, and we measure the robustness of \tratto and the other approaches in terms of the amount of oracles that they can still correctly infer.

\textbf{RQ\textsubscript{5}}: \emph{How effective are the generated oracles for improving test suites?} We apply the oracles that \tratto generates to test suites automatically generated with EvoSuite, and we measure the effectiveness of \tratto for testing as the improvement of the mutation score.

\subsection{RQ\textsubscript{1}: Code Models Comparison}
\label{sec:rq1}

RQ\textsubscript{1} addresses the suitability of LLM code models that \Tratto uses to  both \emph{evaluate oracles} and \emph{predict the next oracle token}.
We compare three models---Code Gemma by
Google~\cite{codegemmateam2024codegemmaopencodemodels}, StarCoder2 by
BigCode~\cite{lozhkov2024starcoder2stackv2}, and Code Llama by
Meta~\cite{rozière2024codellamaopenfoundation}---to determine the most effective LLM for such tasks.
We model both tasks as code infilling problems (see
Listings~\ref{list:oracle-template} and \ref{list:token-template}).  We select these models
since they are state-of-the-art, support infilling problems, are trained on
large-scale code corpora, and are popular among researchers and
practitioners for addressing code-related tasks.
We used the version of each model that features 7B trainable parameters.

The dataset used for training and validation features 223,149 samples (34,249 oracles + 188,900 tokens). We set aside all samples belonging to a single Java project (Guava, 30,891 samples, 14\% of the total) for validation, and used the remaining 192,258 samples (86\%) for training.\footnote{We split the dataset in this way to avoid data leakage, since some oracle and token samples from the same project are the same except for the rephrased Javadoc tag, according to the data augmentation process described in Section~\ref{sec:oracles-dataset}.}
We trained all models for two epochs, as we experimentally determined that accuracy does not improve significantly after this point. We set the input length to 2,048 transformer tokens and the output length to 32 transformer tokens.\footnote{Although the output of the models is a single \emph{lexical} token (e.g., a variable name such as \codeid{millis2secs}) this is tokenized into multiple \emph{transformer} tokens (e.g., \codeid{mill-is-2-se-cs}, five tokens).} We used default hyper-parameters for all trainings.

The accuracy obtained with Code Gemma, StarCoder2, and Code Llama is 58\%, 69\%, and 91\%, respectively. 
Code Llama largely outperforms both Code Gemma and StarCoder2.
We hypothesize that such a difference is due to both the pre-training data and the specificity of each training procedure.
Code Llama is pre-trained primarily on natural language and actual code, and is further specialized for long-context inputs, which fits well with our tasks.
StarCoder2 is pre-trained on varied data, including GitHub issues, pull requests and Jupyter notebooks.
Code Gemma is pre-trained on English language data from open source mathematics datasets and synthetically generated code.
Both Code Gemma and StarCoder2 are pre-trained on general data and synthetic code that are not particularly useful for our case.  

The results of RQ\textsubscript{1} measure the accuracy of different LLM code models in correctly evaluating oracles and selecting oracle tokens, and clearly point to Code Llama as the best choice to generate correct axiomatic oracles.
When used to generate a complete sequence of tokens conforming an oracle, 
\tratto instantiated with Code Llama correctly generates 69\% axiomatic oracles on the Guava validation set, that is, 3,337~oracles (out of 4,865) have all their tokens correctly predicted.\footnote{Note that this dataset also contains non-oracle samples (\emph{Comments without oracles} in Figure~\ref{fig:data-collection}), which are deemed as \emph{correctly predicted} if the \evaluator judged no need to generate an oracle.}

\vspace{-0.1cm}
\begin{tcolorbox}[boxsep=2pt, left=2pt, right=2pt, top=2pt, bottom=2pt]
\textbf{Answer to RQ\textsubscript{1}}: Code Llama achieves
91\% %
accuracy for the tasks of evaluating oracles and selecting tokens, and  outperforms both Code Gemma (58\%) and StarCoder2 (69\%).
\end{tcolorbox}

\subsection{RQ\textsubscript{2}: Ablation Studies}
\label{sec:rq2}

RQ\textsubscript{2} addresses the contributions of the symbolic module and multitask model of  \Tratto in generating oracles.
To this end, we carry out two ablation studies where these components are removed.

In the first study, we remove the token-by-token oracle generation approach enabled by the symbolic module. Thus, the neural module either infers the whole oracle or indicates the impossibility of generating one, solely based on the unit under test. 
We set aside the Guava oracles (4,865) for validation, and used the remaining (29,384) for fine-tuning the Code Llama 7B model with default hyper-parameters for two epochs, as we did for \Tratto.
The resulting model generates 3,049 correct oracles out of 4,865, achieving an accuracy of 63\%, 6\% less that the 69\% accuracy  of \tratto.
The difference is statistically significant according to the McNemar test ($p$-value $<$ 0.001, Odds Ratio = 2.57), which is suitable to pairwise compare dichotomous results of two different treatments as in this case~\cite{mcnemar}.
This indicates that the symbolic module and the token-by-token approach contribute to improve the performance of generating axiomatic oracles.

In the second study, we remove the multitask model and fine-tune two separate models,  an \evaluator and a \selector.
Both are based on Code Llama 7B and the training is performed with default hyper-parameters for two epochs. The \evaluator is fine-tuned on the 29,384 oracle samples (86\% of all oracles) that do not include the Guava project, and the \selector is fine-tuned on the respective 162,874 token samples (86\% of all tokens). Then, we jointly evaluate the capabilities of both models to correctly predict the 4,865 oracles from the Guava validation set, i.e., with the \evaluator correctly predicting whether an oracle should be generated and, if so, with the \selector correctly predicting all tokens of the oracle. Overall, this approach achieved 66\% accuracy in generating oracles (3,213 out of 4,865 oracles from the Guava validation set), 3\% less than the accuracy of \tratto.
The difference is statistically significant ($p$-value $<$ 0.001, Odds Ratio = 1.55).
This highlights the benefits of leveraging a multitask model for evaluating oracles and predicting tokens.

Our analysis of the oracles that \tratto generates and the ablated approaches fail to generate indicates that the ablated approaches either incur in false positives, i.e., they generate oracles where there should be none, or they generate incorrect oracles.
As an example of false positives, the purely neural model wrongly generates the precondition ``\codeid{array != null}'' from the Javadoc tag \emph{``@param array an array of \{@code short\} values, possibly empty''}. The \evaluator of the non-multitask model generates a precondition 
from the Javadoc tag \emph{``@param defaultValue the value provided for inputs absent in map keys''}, which does not express any precondition.
As an example of wrong oracles, the purely neural model generates the incorrect oracle ``\codeid{true ? Arrays.stream(array).anyMatch(jdVar -> jdVar == target) : true}'', which asserts that the \codeid{array} parameter should always contain the \codeid{target} parameter, from the Javadoc tag \emph{``@return \{@code true\} when any element \{@code array[i]\} equals \{@code target\}''}, while \Tratto generates the correct oracle, ``\codeid{Arrays.stream(array).anyMatch(jdVar -> jdVar == target) ? methodResultID == true : methodResultID == false}'', which asserts that, if the \codeid{array} parameter contains the \codeid{target} parameter, the method should return \codeid{true}, otherwise it should return \codeid{false}.

\vspace{-0.2cm}
\begin{tcolorbox}[boxsep=2pt, left=2pt, right=2pt, top=2pt, bottom=2pt]
\textbf{Answer to RQ\textsubscript{2}}: 
The purely neural and the non-multitask approaches achieve 63\% and 66\% accuracy, respectively, less than the 69\% accuracy of \tratto. Thus, we conclude that the symbolic module and the multitask model both improve the performance of \tratto.

\end{tcolorbox}

\subsection{RQ\textsubscript{3}: Oracle Generation}
\label{sec:rq3}

RQ\textsubscript{3} compares the effectiveness of \Tratto to the state-of-the-art neural and symbolic approaches for generating axiomatic oracles. We present the ground-truth dataset used for evaluating, the state-of-the-art approaches we compare \tratto with, the metrics we computed with the experiments, the training and evaluation setup of the experiments, and finally we discuss the results.

\subsubsection{Ground-Truth Dataset}
\label{sec:evaluation-dataset}

We manually created a ground-truth dataset of axiomatic test oracles, to fairly compare \tratto with state-of-the-art approaches. 
We extracted and validated oracles from Defects4J~\cite{Just:Defects4j:ISSTA:2014}. We excluded two projects (Apache Commons Math~\cite{apache_commons_math} and Apache Commons
Collections~\cite{apache_commons_collections}) that are part of the
training set of \Tratto. For each of the remaining 15 projects in
Defects4J, we systematically selected 10 Java classes by sorting them according to their character count
and selecting those evenly spaced within the 5\% to 95\% range (i.e.,
the classes at the percentiles 5, 15, 25, etc.). We manually extracted all
possible axiomatic oracles from all methods of each selected class.
The dataset contains 389 axiomatic oracles (\emph{positive samples}) belonging to 274 methods of 150
classes from 15 projects.
For each method, we also generate one \emph{negative sample} if no precondition, postcondition or exceptional postcondition could be generated for such method. For instance, if a method encodes two preconditions, it will result in two positive samples (two preconditions) plus two negative samples (one \emph{non-postcondition} and one \emph{non-exceptional-postcondition}). This leads to a total of 496 negative samples.
Each sample was reviewed by at least two authors.

\subsubsection{Comparison Approaches}
\label{sec:evaluation-baselines}
We compare \Tratto with Jdoctor~\cite{Blasi:Jdoctor:ISSTA:2018} and GPT4~\cite{gpt4}, as representative symbolic and neural approaches, respectively. 
Jdoctor~\cite{Blasi:Jdoctor:ISSTA:2018} generates axiomatic oracles (preconditions, normal and exceptional postconditions) from the Javadoc tags \codeid{@param}, \codeid{@return} and \codeid{@throws}/\codeid{@exception}, respectively, by applying pattern, lexical, and semantic matching techniques. Jdoctor outperforms other approaches such as Toradocu~\cite{Goffi:Exceptional:ISSTA:2016} and @tComment~\cite{tan:tcomment:ICST:2012}.

The state-of-the-art neural approaches to generate oracles~\cite{Tufano:Transformers:AST:2022,Watson:AssertionGenerationNN:ICSE:2020,Dinella:TOGA:ICSE:2022} produce concrete test oracles, that is, assertions on specific concrete inputs. They are not directly comparable with \tratto, which generates axiomatic oracles that predicate on variables, and are valid for all concrete inputs.   
Thus we compare \tratto 
against GPT4 (model GPT4-o) as representative neural approach to generate axiomatic oracles. We enhance GPT4 with few shot learning~\cite{Wang:few-shot-learning:csur:2022} combined with Chain-of-Thought
prompting~\cite{Wei:Chain-of-Thought:2022} to fully leverage its understanding capabilities. 
We provide GPT4 with three examples of
how and when to generate axiomatic oracles based on several Java methods,
following a step-by-step approach, and ask it to generate oracles for the
ground-truth dataset, method by method. We share the prompts we use in our experiments in the replication package~\cite{supplementarymaterial}.

\subsubsection{Metrics}
\label{sec:evaluation-metrics}
We comparatively evaluate \tratto by computing \emph{true positives} (correctly predicted oracles), \emph{true negatives} (instances for which no oracle should be generated, and none is generated), \emph{false positives} (instances for which no oracle should be generated, but one is generated, or wrongly generated oracles), and \emph{false negatives} (oracles not generated, while one should be generated).

We classify incomplete oracles, that is, oracles that capture only a subset of the correct behavior, as \textit{false positives}, since they partially miss the semantic of the reference oracle and do not work for all test cases. The oracle
``\codeid{result != null}'' is an example of \emph{incomplete} oracle for a method that returns a positive \codeid{Integer}, as it correctly captures only a subset of the expected behavior, while ``\codeid{result >= 0}'' is a \emph{wrong} oracle for the same method, as 0 is not positive.

We also compute \textit{accuracy} (portion of all correct predictions out of the total predictions), \emph{precision} (portion of oracles correctly generated out of all oracles generated), \emph{recall} (portion of oracles correctly generated out of all actual oracles in the dataset), and \emph{F1-score} (weighted harmonic mean of precision and recall).
\textit{Accuracy} measures the overall performance of the technique, but does not differentiate between \emph{false positives} and \emph{negatives}. \emph{Precision} measures the ability to avoid false positives.  
\emph{Recall} reflects the ability to capture all relevant instances, avoiding false negatives. 
\emph{F1-score} combines  \emph{precision} and \emph{recall} into a single measure that balances false positives and negatives, and is especially useful when the dataset is unbalanced and the positive class is rare, as in our case.

\subsubsection{Training and Evaluation Setup}
\label{sec:evaluation-training-testing}

We trained \Tratto with the same setup as in previous experiments, although using the complete dataset (223k samples) for training. We generated  as many axiomatic oracles as possible for all methods in the ground-truth dataset.
We downloaded and set up the most recent version of Jdoctor from the publicly available GitHub repository (commit \href{https://github.com/albertogoffi/toradocu/commit/d76899fc65bf0c8912479606d61998d91f30e5bd}{
d76899f}). Jdoctor does not need training, as it is based on a set of heuristic rules and patterns.  We generated  as many axiomatic oracles as possible for all methods in the ground-truth dataset, as we did for \tratto. 
We performed similarly for GPT4, generating as many axiomatic oracles as possible for all methods,
by crafting a prompt per method, making an API call per prompt, and collecting all responses, which we manually analyzed.

\subsubsection{Results}
\label{sec:results}

Table~\ref{tab:apr} reports the \emph{accuracy} (row $A$), \emph{precision} ($P$),  \emph{recall} ($R$), and \emph{F1-score} ($F1$) that we computed for each approach (\tratto, Jdoctor and GPT4) and each project (columns \textit{closure-compiler} ... \textit{mockito}), as well as for all projects (column \textit{Total}). 
The table highlights in green the best \emph{Total} values and in red the worst values.
We observe that \tratto outperforms both Jdoctor and GPT4 in terms of \emph{accuracy} (73\% for \tratto, 61\% for Jdoctor, 40\% for GPT4), \emph{precision} (72\% for \tratto, 62\% for Jdoctor,  24\% for GPT4), and \emph{F1-score} (61\% for \tratto, 25\% for Jdoctor, 37\% for GPT4).  \tratto's \emph{recall} (52\%) is better than Jdoctor's (16\%) and worse than GPT4 (89\%).
The results indicate that \tratto infers more correct predictions than both Jdoctor and GPT4 (\emph{accuracy}) with a low impact of wrong results (\emph{precision}).  The better performance of GPT4 than \tratto in terms of \emph{recall} indicates that GPT4 generates a higher proportion of oracles than \tratto out of the ground truth.
This was expected since GPT4 is a general-purpose LLM with better generalization capabilities. However, the higher recall comes with a cost of more false positives (lower precision), on which \tratto does not incur as often thanks to the fine-tuning performed, which allows it to evaluate more precisely whether an oracle should be generated or not, and what the shape or the oracle should be. This is also illustrated with the \emph{F1-score}, which well summarizes the improvement of the token-by-token neuro-symbolic approach of \tratto over the symbolic approach of Jdoctor and the neural approach of GPT4. 
The precision and recall of the approaches vary greatly across projects, thus confirming the dependency of the approaches on the quality of comments and code. 

Table~\ref{tab:results} provides futher details for each project and approach. The table reports the number of methods (column $M$), the number of oracles in the ground truth ($O$), the number of non-oracle (negative) instances ($NO$), and the true and false predictions ($TP$, $TN$, $FP$, $FN$) for all approaches and projects, as well as the totals (last row).
For each metric, we report both the total number of predictions as well as the percentage over the total number of instances ($O + NO = 389 + 496 = 885$).
Note that the sum of $TP + TN + FP + FN$ may go over 885 as any approach can generate any arbitrary number of false positives.

\begin{table}[]
\centering
\caption{Accuracy ($A$), precision ($P$), recall ($R$), and F1-score ($F1$) for all approaches on ground-truth dataset.}
\vspace{-0.5cm}
\scriptsize %
\begin{adjustbox}{width=\textwidth} %
\begin{tabular}{cc|*{16}{>{\raggedleft\arraybackslash}b{0.5cm}}c}
\multicolumn{2}{c}{}                       & 
\makecell[rb]{\rotatebox{25}{closure-compiler}} & 
\makecell[rb]{\rotatebox{25}{commons-cli}} & 
\makecell[rb]{\rotatebox{25}{commons-codec}} & 
\makecell[rb]{\rotatebox{25}{commons-compress}} & 
\makecell[rb]{\rotatebox{25}{commons-csv}} & 
\makecell[rb]{\rotatebox{25}{commons-jxpath}} & 
\makecell[rb]{\rotatebox{25}{commons-lang}} & 
\makecell[rb]{\rotatebox{25}{gson}} & 
\makecell[rb]{\rotatebox{25}{jackson-core}} & 
\makecell[rb]{\rotatebox{25}{jackson-databind}} & 
\makecell[rb]{\rotatebox{25}{jackson-dataformat}} & 
\makecell[rb]{\rotatebox{25}{jfree-chart}} & 
\makecell[rb]{\rotatebox{25}{joda-time}} & 
\makecell[rb]{\rotatebox{25}{jsoup}} & 
\makecell[rb]{\rotatebox{25}{mockito}} & 
\makecell[rb]{\rotatebox{25}{Total}} \\ \hline
\multicolumn{1}{c|}{}                          & A & 56\%                                 & 83\%                            & 61\%                              & 74\%                                 & 91\%                            & 62\%                               & 75\%                             & 80\%                     & 68\%                             & 58\%                                 & 67\%                                       & 77\%                            & 81\%                          & 60\%                      & 67\%                         & \multicolumn{1}{|c}{\color[HTML]{036400} \textbf{73\%}} \\
\multicolumn{1}{c|}{}                          & P & N/A                                  & 60\%                            & 29\%                              & 82\%                                 & 88\%                            & N/A                                & 61\%                             & 70\%                     & 43\%                             & 33\%                                 & N/A                                        & 90\%                            & 83\%                          & 50\%                      & N/A                          & \multicolumn{1}{|c}{\color[HTML]{036400} \textbf{72\%}} \\
\multicolumn{1}{c|}{}  & R & 0\%                                  & 75\%                            & 24\%                              & 53\%                                 & 97\%                            & 0\%                                & 67\%                             & 70\%                     & 33\%                             & 3\%                                  & 0\%                                        & 60\%                            & 78\%                          & 5\%                       & 0\%                          & \multicolumn{1}{|c}{52\%}                        \\
\multicolumn{1}{c|}{\multirow{-4}{*}{\tratto}}  & F1 & N/A                                  & 67\%                            & 26\%                              & 64\%                                 & 92\%                            & N/A                                & 64\%                             & 70\%                     & 37\%                             & 6\%                                  & N/A                                        & 72\%                            & 80\%                          & 8\%                       & N/A                          & \multicolumn{1}{|c}{\color[HTML]{036400} \textbf{61\%}}                        \\ \hline
\multicolumn{1}{c|}{}                          & A & 56\%                                 & 72\%                            & 56\%                              & 73\%                                 & 79\%                            & 62\%                               & 60\%                             & 64\%                     & 67\%                             & 61\%                                 & 67\%                                       & 47\%                            & 59\%                          & 57\%                      & 67\%                         & \multicolumn{1}{|c}{61\%}                        \\
\multicolumn{1}{c|}{}                          & P & N/A                                  & 100\%                           & 33\%                              & 100\%                                & 96\%                            & N/A                                & 0\%                              & 6\%                      & 33\%                             & 100\%                                & N/A                                        & N/A                             & 74\%                          & 0\%                       & N/A                          & \multicolumn{1}{|c}{62\%}                        \\
\multicolumn{1}{c|}{} & R & 0\%                                  & 16\%                            & 4\%                               & 41\%                                 & 68\%                            & 0\%                                & 0\%                              & 7\%                      & 11\%                             & 3\%                                  & 0\%                                        & 0\%                             & 33\%                          & 0\%                       & 0\%                          & \multicolumn{1}{|c}{\color[HTML]{CB0000} \textbf{16\%}} \\
\multicolumn{1}{c|}{\multirow{-4}{*}{Jdoctor}} & F1 & N/A                                  & 29\%                            & 7\%                               & 58\%                                 & 79\%                            & N/A                                & N/A                              & 6\%                      & 17\%                             & 6\%                                  & N/A                                        & N/A                             & 46\%                          & N/A                       & N/A                          & \multicolumn{1}{|c}{\color[HTML]{CB0000} \textbf{25\%}} \\ \hline
\multicolumn{1}{c|}{}                          & A & 18\%                                 & 26\%                            & 37\%                              & 42\%                                 & 55\%                            & 5\%                                & 53\%                             & 19\%                     & 47\%                             & 35\%                                 & 50\%                                       & 35\%                            & 53\%                          & 41\%                      & 50\%                         & \multicolumn{1}{|c}{\color[HTML]{CB0000} \textbf{40\%}} \\
\multicolumn{1}{c|}{}                          & P & 12\%                                 & 7\%                             & 20\%                              & 27\%                                 & 40\%                            & 2\%                                & 31\%                             & 3\%                      & 22\%                             & 13\%                                 & 0\%                                        & 27\%                            & 42\%                          & 23\%                      & 33\%                         & \multicolumn{1}{|c}{\color[HTML]{CB0000} \textbf{24\%}} \\
\multicolumn{1}{c|}{}  & R & 100\%                                & 100\%                           & 92\%                              & 100\%                                & 92\%                            & 100\%                              & 88\%                             & 60\%                     & 100\%                            & 100\%                                & N/A                                        & 92\%                            & 96\%                          & 67\%                      & 100\%                        & \multicolumn{1}{|c}{\color[HTML]{036400} \textbf{89\%}} \\
\multicolumn{1}{c|}{\multirow{-4}{*}{GPT4}}  & F1 & 22\%                                & 12\%                           & 32\%                              & 42\%                                & 56\%                            & 4\%                              & 46\%                             & 6\%                     & 36\%                            & 24\%                                & N/A                                        & 42\%                            & 58\%                          & 34\%                      & 50\%                        & \multicolumn{1}{|c}{37\%} \\\hline
\end{tabular}
\end{adjustbox}
\label{tab:apr}
\vspace{-0.5cm}
\end{table}

Row \emph{Total} of Table~\ref{tab:results} highlights in green the best and in red the worst performance of the three approaches.  \tratto generates the highest rate of correct oracles (total true positive rate 21\%), and avoids generating oracles that shall not be generated in a good number of cases (total true negative rate 52\%, just below the best result of Jdoctor, 54\%).  \tratto performs well also in terms of false alarms and missed oracles, with low false positive/negative rates, i.e., 8\% and 19\%, respectively and overall.  Jdoctor performs sightly better than \tratto in terms of true negative (54\% vs. 52\%) and false positive rates (4\% vs. 8\%), however, it performs worst among the three approaches in terms of both true positive (7\%) and false negative rates (35\%).  GPT4 presents an excellent false negative rate (only 2\%) however with poor true positive (18\%), true negative (22\%) and false positive rates (58\%). 
The true and false positive and negative rates confirm the best performance of \tratto among the three approaches, as well summarized by the best \emph{F1-score} in Table~\ref{tab:apr}.

Overall, \Tratto generates 186 out of 389 oracles in the ground truth and demonstrates a good capability to discern when an oracle should be generated, with a relatively small number of wrong oracles (72 out of 885 predictions).  Thus, \Tratto can significantly reduce the effort of manually generating oracles without a big overhead for identifying and discarding wrong oracles.

\Tratto generates three times more correct oracles than Jdoctor (186 vs. 58).  The result is not surprising: Jdoctor exploits classic natural language processing and semantic matching, and as such, it works well in the presence of precise comments in natural language, however it does not process well the imprecise comments that often occur in Javadoc documentation.  The neuro-symbolic approach of \Tratto is much more tolerant with respect to the precision of the comments in natural language, and handles many more cases than Jdoctor. The results highly depend on the quality of the comments and code. Jdoctor generates a fair number of oracles for \codeid{commons-csv} (23 oracles while \Tratto generates 30), but does not generate any oracle for \codeid{jfree-chart}, while \Tratto generates 26 correct oracles. 
GPT4 infers 217 correct oracles,   
a relative small increment with respect to \Tratto (186 correct oracles). This reflects the huge difference in the size of the respective neural components (GPT4 features hundreds of billions of parameters against the 7 billion parameters of Code Llama, the reference model of \Tratto). 
GPT4 generates 701 false positives, almost 10 times more than \tratto (72), which accounts for more than half of all the predictions (58\%).  
The significant amount of false positives showcases the main difference between the pure neural approach of GPT4 and the neuro-symbolic approach of \tratto: the symbolic component of \tratto discriminates valid from invalid results, while its neural component is specifically trained for discerning the possibility to generate and oracle (\evaluator), thus \tratto generates less valid results than GPT4 but also a much more limited number of invalid results, while GPT4 does not check the validity and compilability of the results, and often generates oracles in cases where none should be generated.
The large amount of false positives of GPT4 greatly reduces its practical applicability since it requires massive human effort to prune the results.

\begin{table}[]
\setlength\extrarowheight{2pt}
\centering
\caption{True/false positives/negatives ($TP$/$TN$/$FP$/$FN$) for all approaches on ground-truth dataset.}
\vspace{-0.25cm}
\scriptsize %
\begin{adjustbox}{width=\textwidth} %
\begin{tabular}{m{0.144\textwidth}W{0.015\textwidth}W{0.015\textwidth}W{0.015\textwidth}|G{0.007\textwidth}G{0.023\textwidth}G{0.007\textwidth}G{0.023\textwidth}W{0.007\textwidth}W{0.023\textwidth}W{0.007\textwidth}W{0.023\textwidth}|G{0.007\textwidth}G{0.023\textwidth}G{0.007\textwidth}G{0.023\textwidth}W{0.007\textwidth}W{0.023\textwidth}W{0.007\textwidth}W{0.023\textwidth}|G{0.007\textwidth}G{0.023\textwidth}G{0.007\textwidth}G{0.023\textwidth}W{0.007\textwidth}W{0.023\textwidth}W{0.007\textwidth}W{0.023\textwidth}}
\hline
\multicolumn{1}{c}{\multirow{2}{*}{Project}} & \multirow{2}{*}{$M$} & \multirow{2}{*}{O} & \multirow{2}{*}{$NO$} & \multicolumn{8}{c|}{\tratto}                              & \multicolumn{8}{c|}{Jdoctor}                            & \multicolumn{8}{c}{GPT4}                               \\ \cline{5-28}
  \multicolumn{4}{c|}{} & $TP$  &        & $TN$  &        & $FP$ &        & $FN$  &        & $TP$ &        & $TN$  &        & $FP$ &        & $FN$  &        & $TP$  &        & $TN$  &        & $FP$  &        & $FN$ &        \\ \hline
closure-compiler                        & 3           & 4          & 5        & 0   & (0\%)  & 5   & (56\%) & 0  & (0\%)  & 4   & (44\%) & 0  & (0\%)  & 5   & (56\%) & 0  & (0\%)  & 4   & (44\%) & 2   & (12\%) & 1   & (6\%)  & 14  & (0\%)  & 0  & (82\%) \\
commons-cli                             & 6           & 6          & 12        & 3   & (16\%) & 12  & (67\%) & 2  & (6\%)  & 1   & (11\%) & 1  & (5\%)  & 12  & (67\%) & 0  & (0\%)  & 5   & (28\%) & 1   & (5\%)  & 4   & (21\%) & 14  & (74\%) & 0  & (0\%)  \\
commons-codec                          & 19           & 24         & 33        & 4   & (7\%)  & 32  & (54\%) & 10 & (22\%) & 13  & (17\%) & 1  & (2\%)  & 31  & (54\%) & 2  & (4\%)  & 23  & (40\%) & 11  & (15\%) & 16  & (22\%) & 45  & (62\%) & 1  & (1\%)  \\
commons-compress                      & 12           & 17          & 20       & 9   & (24\%) & 19  & (50\%) & 2  & (5\%)  & 8   & (21\%) & 7  & (19\%) & 20  & (54\%) & 0  & (0\%)  & 10  & (27\%) & 12  & (21\%) & 12  & (21\%) & 33  & (58\%) & 0  & (0\%)  \\
commons-csv                            & 16           & 35         & 23        & 30  & (52\%) & 23  & (40\%) & 4  & (7\%)  & 1   & (2\%)  & 23 & (39\%) & 23  & (39\%) & 1  & (2\%)  & 11  & (18\%) & 24  & (28\%) & 23  & (27\%) & 36  & (42\%) & 2  & (2\%)  \\
commons-jxpath                          & 4           & 5          & 8        & 0   & (0\%)  & 8   & (62\%) & 0  & (0\%)  & 5   & (38\%) & 0  & (0\%)  & 8   & (62\%) & 0  & (0\%)  & 5   & (38\%) & 1   & (2\%)  & 2   & (4\%)  & 49  & (94\%) & 0  & (0\%)  \\
commons-lang                           & 53           & 64         & 103        & 37  & (22\%) & 89  & (53\%) & 24 & (14\%) & 18  & (11\%) & 0  & (0\%)  & 101 & (61\%) & 2  & (1\%)  & 64  & (38\%) & 38  & (20\%) & 62  & (33\%) & 83  & (44\%) & 5  & (3\%)  \\
gson                                  & 27           & 30          & 51       & 19  & (23\%) & 46  & (57\%) & 8  & (10\%) & 8   & (10\%) & 1  & (1\%)  & 51  & (63\%) & 16 & (20\%) & 13  & (16\%) & 3   & (3\%)  & 18  & (16\%) & 88  & (79\%) & 2  & (2\%)  \\
jackson-core                           & 10           & 10         & 20        & 3   & (10\%) & 18  & (58\%) & 4  & (13\%) & 6   & (19\%) & 1  & (3\%)  & 19  & (63\%) & 2  & (7\%)  & 8   & (27\%) & 5   & (15\%) & 11  & (32\%) & 18  & (53\%) & 0  & (0\%)  \\
jackson-databind                       & 24           & 30         & 48        & 1   & (1\%)  & 42  & (57\%) & 2  & (3\%)  & 29  & (39\%) & 1  & (1\%)  & 46  & (60\%) & 0  & (0\%)  & 29  & (39\%) & 13  & (10\%) & 33  & (25\%) & 84  & (65\%) & 0  & (0\%)  \\
jackson-dataformat                      & 1           & 1          & 2        & 0   & (0\%)  & 2   & (67\%) & 0  & (0\%)  & 1   & (33\%) & 0  & (0\%)  & 2   & (67\%) & 0  & (0\%)  & 1   & (33\%) & 0   & (0\%)  & 2   & (50\%) & 2   & (50\%) & 0  & (0\%)  \\
jfree-chart                            & 25           & 46         & 40        & 26  & (0\%)  & 40  & (47\%) & 3  & (3\%)  & 17  & (20\%) & 0  & (0\%)  & 40  & (47\%) & 0  & (0\%)  & 46  & (53\%) & 34  & (24\%) & 17  & (12\%) & 90  & (62\%) & 3  & (2\%)  \\
joda-time                              & 40           & 71         & 63        & 52  & (39\%) & 56  & (42\%) & 11 & (8\%)  & 15  & (11\%) & 23 & (17\%) & 56  & (42\%) & 8  & (6\%)  & 47  & (35\%) & 50  & (33\%) & 31  & (20\%) & 70  & (46\%) & 2  & (1\%)  \\
jsoup                                  & 33           & 45         & 66        & 2   & (2\%)  & 65  & (59\%) & 2  & (2\%)  & 42  & (38\%) & 0  & (0\%)  & 64  & (57\%) & 4  & (3\%)  & 45  & (40\%) & 22  & (16\%) & 36  & (25\%) & 73  & (51\%) & 11 & (8\%)  \\
mockito                                & 1            & 1          & 2        & 0   & (0\%)  & 2   & (67\%) & 0  & (0\%)  & 1   & (33\%) & 0  & (0\%)  & 2   & (67\%) & 0  & (0\%)  & 1   & (33\%) & 1   & (25\%) & 1   & (25\%) & 2   & (50\%) & 0  & (0\%)  \\ \hline
Total                  
                                      & 274           & 389        & 496        & 186 & ({\color[HTML]{036400}\textbf{21\%}}) & 459 & (52\%) & 72 & (8\%) & 169 & (19\%) & 58 & ({\color[HTML]{CB0000}\textbf{7\%}})  & 480 & ({\color[HTML]{036400}\textbf{54\%}}) & 35 & ({\color[HTML]{036400}\textbf{4\%}}) & 312 & ({\color[HTML]{CB0000}\textbf{35\%}})  & 217 & (18\%) & 269 & ({\color[HTML]{CB0000}\textbf{22\%}}) & 701 & ({\color[HTML]{CB0000}\textbf{58\%}}) & 26 & ({\color[HTML]{036400}\textbf{2\%}})  \\ \hline
\end{tabular}
\end{adjustbox}
\label{tab:results}
\vspace{-0.5cm}
\end{table}

GPT4 produces very few false negatives (26, 2\% of the total predictions), while \Tratto and Jdoctor do not generate any oracles for 169 and 312 oracles in the ground truth (19\% and 35\% of the total predictions), respectively. 
The good performance of GPT4 in terms of false negatives balances the bad results in terms of false positives: GPT4 mostly always attempts to generate oracles. As a consequence, it generates lots of wrong oracles (false positives) and misses very few cases (false negatives). On the contrary, both \tratto and Jdoctor identify those cases where no oracle should be generated, and generate less wrong oracles.
The overall performance of the approaches is well reflected by the combination of false predictions (positives and negatives): GPT4 generates a total of 727 false results (701 FP + 26 FN), Jdoctor 347 (35 + 312) and \tratto only 241 (72 + 169).

The second \codeid{@throws} tag in Listing~\ref{list:tratto-except-example} (line 6) well exemplifies the capability of \tratto to infer correct oracles from Javadoc tags in terms of exceptional postconditions that both Jdoctor and GPT4 fail to generate. The \codeid{@throws} tag contains an implicit reference (line 6, \emph{``this method is called on a closed result set''}) to a method (\codeid{isClosed}) of a class (\codeid{ResultSet}) of an external library (\codeid{java.sql}): \tratto's symbolic module retrieves the contextual information related to the external class from the signature of the method (line 9, \codeid{resultSet} method parameter) and feeds the neural model with additional information to produce the correct axiomatic oracle ``\codeid{resultSet.isClosed();}''. Jdoctor cannot infer an oracle from the Javadoc tag, since it falls outside the set of rules and patterns, resulting in a \emph{false negative}. GPT4 generates the precondition ``\codeid{resultSet != null;}'', claiming that the parameter \codeid{resultSet} must not be null, even if neither the documentation nor the contextual information indicate that the method cannot accept null values in input. GPT4 generates also three exceptional postconditions (``\codeid{resultSet == null ? NullPointerException;}'', ``\codeid{IOException may be thrown;}'' and ``\codeid{SQLException may be thrown if resultSet is closed or if there is a database access error}'') that are either wrong or not compilable. The example highlights how a pure neural model struggles to infer non-trivial oracles, may generate  non-compilable oracles since it lacks a precise grammar, and generates a non-negligible number of \emph{false positives}.

A manual inspection of the oracles generated with \tratto and Jdoctor shows that there is only one oracle that Jdoctor generates and \tratto does not. This was an exceptional oracle stating the need of a method parameter to be of a certain type (via the \codeid{instanceof} operator). We hypothesize that \tratto's training set did not include enough oracles of this kind and thus the \evaluator judged no need to generate an oracle in this case. On the other hand, \tratto generates significantly more oracles than Jdoctor, as the symbolic matching of Jdoctor often overlooks similar cases. 
For example, \tratto successfully generates exceptional oracles ``\codeid{source == null;}'' and ``\codeid{bigInteger == null;}'' from the Javadoc comments  \emph{``the parameter passed to this method is null''} in Listing~\ref{list:positive-jdoctor-example} (lines 14-15) and \emph{``@throws NullPointerException if null is passed in''} in Listing~\ref{list:negative-jdoctor-example} (line 8), respectively. 
Jdoctor generates only the first oracle, and cannot generate the second, despite the similarity of the two Javadoc comments. 
The neural component of \tratto enhances \tratto's ability to identify interesting oracles and explains the significantly higher recall of \tratto over Jdoctor.

\begin{lstlisting}[language=java, caption={Documentation and implementation of method \href{https://github.com/apache/commons-csv/blob/2f44a6b399edf1d243b340180538a7f657a2a078/src/main/java/org/apache/commons/csv/CSVPrinter.java\#L232}{printHeaders} from \href{https://github.com/apache/commons-csv/blob/2f44a6b399edf1d243b340180538a7f657a2a078/src/main/java/org/apache/commons/csv/CSVPrinter.java}{CSVPrinter}.},
  label=list:tratto-except-example, numberstyle=\tiny\color{pgrey}]
/**
* Prints headers for a result set based on its metadata.
*
* @param resultSet The ResultSet to query for metadata.
* @throws IOException If an I/O error occurs.
* @throws SQLException If a database access error occurs or this method is called on a closed result set.
* @since 1.9.0
*/
public synchronized void printHeaders(final ResultSet resultSet) throws IOException, SQLException {
  printRecord((Object[]) format.builder().setHeader(resultSet).build().getHeader());
}
\end{lstlisting}

\begin{adjustbox}{width=\textwidth}
\noindent\begin{minipage}[b]{0.515\textwidth}

\begin{lstlisting}[language=java, caption={Documentation and implementation of method \href{https://github.com/apache/commons-codec/blob/3dbd866654140fada4616647efeedaa43d26b5d8/src/main/java/org/apache/commons/codec/Decoder.java\#L45}{decode} from \href{https://github.com/apache/commons-codec/blob/master/src/main/java/org/apache/commons/codec/Decoder.java}{Decoder}.},
  label=list:positive-jdoctor-example, numberstyle=\tiny\color{pgrey}]
/**
* Decodes an "encoded" Object and returns a
* "decoded" Object. Note that the implementation of
* this interface will try to cast the Object
* parameter to the specific type expected by a
* particular Decoder implementation. If {@link
* ClassCastException} occurs this decode method will
* throw a DecoderException.
*
* @param source the object to decode
* @return a 'decoded" object
* @throws DecoderException a decoder exception can
* be thrown for any number of reasons. Some good
* candidates are that the parameter passed to this
* method is null, a param cannot be cast to the
* appropriate type for a specific encoder.
*/
Object decode(Object source) throws DecoderException;
\end{lstlisting}

\end{minipage}
\hspace{0.02\textwidth}
\begin{minipage}[b]{0.51\textwidth} 

\begin{lstlisting}[language=java, caption={Documentation and implementation of method \href{https://github.com/apache/commons-codec/blob/3dbd866654140fada4616647efeedaa43d26b5d8/src/main/java/org/apache/commons/codec/binary/Base64.java\#L376}{encodeInteger} from \href{https://github.com/apache/commons-codec/blob/3dbd866654140fada4616647efeedaa43d26b5d8/src/main/java/org/apache/commons/codec/binary/Base64.java}{Base64}.},
  label=list:negative-jdoctor-example, numberstyle=\tiny\color{pgrey}]
/**
* Encodes to a byte64-encoded integer according to
* crypto standards such as W3C's XML-Signature.
*
* @param bigInteger a BigInteger
* @return A byte array containing base64 character
* data
* @throws NullPointerException if null is passed in
* @since 1.4
*/
public static byte[] encodeInteger(final BigInteger
      bigInteger) {
  Objects.requireNonNull(bigInteger, "bigInteger");
  return encodeBase64(toIntegerBytes(bigInteger),
          false);
}
\end{lstlisting}

\end{minipage}
\end{adjustbox}

\vspace{-0.2cm}
\begin{tcolorbox}[boxsep=2pt, left=2pt, right=2pt, top=2pt, bottom=2pt]
\textbf{Answer to RQ\textsubscript{3}}: 
\tratto achieves an F1-score of 61\% in generating axiomatic oracles, significantly higher than Jdoctor (25\%) and GPT4 (37\%). \tratto can generate a high number of oracles (186, $3\times$ more than Jdoctor) incurring in few false positives (72, $10\times$ less than GPT4).
\end{tcolorbox}

\subsection{RQ\textsubscript{4}: Robustness to Documentation Variations}
\label{sec:rq4}
RQ\textsubscript{4} evaluates the robustness of \tratto, Jdoctor and GPT4 in terms of their ability to generate axiomatic oracles with varying degrees of documentation quality. We select all oracles that all three approaches successfully generate (RQ\textsubscript{3}) and systematically generate variations of their associated documentation. We then compute the proportion of oracles that the approaches can still generate with the modified documentation.

There are 55 oracles that all three approaches successfully generate from eight projects from the ground-truth dataset. For each oracle, we identify the Javadoc tag based on which the oracle was generated, and we generate four variations of it. We automate the process and make it systematic, by asking ChatGPT to generate alternative versions of the Javadoc tag with the following criteria: (i)~replacing words with synonyms or rephrasing sentences while keeping the same meaning; (ii)~changing the order of the sentence; (iii)~introducing grammatical mistakes or typos; and (iv)~making the sentence less explicit. Listing~\ref{list:javadoc-tags-rephrased} shows some examples of the generated variations.

\begin{wraptable}{r}{0.6\textwidth}
\centering
\caption{Robustness to documentation variations.}
\vspace{-0.25cm}
\begin{adjustbox}{width=0.6\textwidth}
\begin{tabular}{lrrrrr}
\toprule
Approach & Synonyms & Order & Typos & Explicitness & Total \\
\midrule
\tratto  & 53 (96\%) & 54 (98\%) & 54 (98\%) & 34 (62\%) & 195 (89\%) \\
Jdoctor & 16 (29\%) & 29 (53\%) & 4 (7\%) & 0 (0\%) & 49 (22\%) \\
GPT4    & 52 (95\%) & 54 (98\%) & 53 (96\%) & 49 (89\%) & 208 (95\%) \\
\bottomrule
\end{tabular}
\end{adjustbox}
\label{tab:robustness-evaluation}
\end{wraptable}

We manually check the correctness of the oracles that the approaches generate for each of the 220 variations (55 oracles $\times$ 4 variations). Table~\ref{tab:robustness-evaluation} shows the number and percentage of oracles that each approach correctly generates. \tratto generates about 90\% of correct oracles for the documentation variations, about the same as GPT4, despite relying on a much smaller model. Indeed, the performance of \tratto is similar to GPT4 for all types of variations except for less explicit descriptions. This is expected since these do not explicitly provide information regarding oracles (e.g., line 6 in Listing~\ref{list:javadoc-tags-rephrased}). Jdoctor is the least robust approach, with only 22\% correct oracles for the variations, and no oracles at all for less explicit descriptions.

\begin{lstlisting}[caption={Rephrased Javadoc tags generated with ChatGPT.}, label=list:javadoc-tags-rephrased, numberstyle=\tiny\color{pgrey}]
(object == null) == false; // Generated oracle
@param object  the object to convert, must not be null // Original Javadoc tag
@param object the item to transform, should not be null // Synonyms or rephrasing
@param object Must not be null, this is the object that needs conversion. // Changed order
@param object the obect to convirt, musn't be nul // Grammatical mistakes or typos
@param object an element to be used, should be valid // Less explicit
\end{lstlisting}
\vspace{-0.4cm}

\subsection{RQ\textsubscript{5}: Application to Software Testing}
\label{sec:rq5}
We address RQ\textsubscript{5} with an exploratory study on the impact of the automatically generated oracles on the mutation score~\cite{Jia:Analysis:TSE:2011} of test suites. We developed a script to automatically insert the generated oracles into test cases that contain a call to a method for which an oracle was generated. Oracles that make a test case fail are simply discarded, although a further check on them may discover bug-revealing tests, further strengthening the usefulness of the 
oracles.
Then, the mutation score is computed with the PIT mutation testing tool~\cite{coles2016pit}. We measure the impact of the oracles as the difference between the mutation score of the test suite with and without the oracles.
We excluded GPT4 from this experiment due to the large amount of non-compilable oracles it generates (552 out of 1,213 in RQ\textsubscript{3}), which makes it impossible to automate the insertion of oracles into test suites.

We generated test suites with EvoSuite~\cite{Fraser:EvoSuite:ESECFSE:2011} for the 10 classes of each project from the ground-truth dataset for which we generated oracles as part of RQ\textsubscript{3}. We were unable to run EvoSuite on five projects---\codeid{closure-compiler}, \codeid{jackson-core}, \codeid{jackson-databind}, \codeid{jackson-dataformat} and \codeid{mockito}---due to incompatibility issues. We did not consider three projects for which Jdoctor did not generate oracles from the ground-truth dataset---\codeid{commons-lang}, \codeid{jfree-chart} and \codeid{jsoup}---and a project for which neither \tratto nor Jdoctor generated oracles---\codeid{commons-jxpath}---since, without oracles, the test suite and consequently the mutation score do not change. By focusing on the projects for which \emph{both} \tratto and Jdoctor generate oracles, we aim to ensure a fair comparison.

We evaluated the improvement achieved with the automatically generated oracles upon test suites with implicit oracles only, and with both implicit and regression oracles. The latter represent the default test suites generated with EvoSuite, while the former entail a more realistic scenario, where regression oracles are not available, or it cannot be assumed that they are correct. To create these test suites, we simply wrap the assertions generated by EvoSuite into try-catch blocks.

Table~\ref{tab:mutation-score} shows the mutation score of the test suites with Evosuite, \tratto and Jdoctor, with and without regression oracles.
The oracles generated by \tratto improve the mutation score with respect to Evosuite's implicit oracles for 5 out of 6 projects, from 2\% (\codeid{commons-compress}) up to 17\% (\codeid{commons-codec}), with an average increase of 7\% across all projects. The oracles improve the mutation score also in the presence of Evosuite's regression oracles on 3 out of 6 projects with an average increase of 1\%. This lower increase was expected since EvoSuite's regression oracles make concrete assumptions of expected behavior and thus may be more effective than axiomatic oracles in this context, although not so widely applicable nor necessarily correct. Table~\ref{tab:mutation-score} also highlights the worse performance of Jdoctor compared to \tratto, as it improved the mutation score of 4 out of 6 test suites with implicit oracles (1\% average increase vs. 7\% of \tratto) and in no cases for test suites with regression oracles.

\begin{wraptable}{r}{0.55\textwidth}
\vspace{-0.75cm}
\huge
\centering
\caption{Mutation score of test suites.}
\vspace{-0.25cm}
\begin{adjustbox}{width=0.55\textwidth}
\begin{tabular}{lrrrrrr}
\toprule
\multirow{2}{*}{Project} & \multicolumn{3}{c}{Implicit oracles} & \multicolumn{3}{c}{Implicit/regression oracles} \\
\cmidrule(lr){2-4} \cmidrule(lr){5-7}
  & EvoSuite & \tratto & Jdoctor & EvoSuite & \tratto & Jdoctor \\
\midrule
commons-cli       & 21\% & \textbf{28\%} & 22\% & 61\% & 61\% & 61\% \\
commons-codec     & 42\% & \textbf{59\%} & 46\% & 71\% & \textbf{75\%} & 71\% \\
commons-compress  & 20\% & \textbf{22\%} & 20\% & 41\% & 41\% & 41\% \\
commons-csv       & 5\% & 5\% & 5\% & 13\% & 13\% & 13\% \\
gson              & 22\% & \textbf{30\%} & 24\% & 67\% & \textbf{68\%} & 67\% \\
joda-time         & 38\% & \textbf{47\%} & 39\% & 78\% & \textbf{79\%} & 78\% \\
\midrule
Total             & 25\% & \textbf{32\%} & 26\% & 55\% & \textbf{56\%} & 55\% \\
\bottomrule
\end{tabular}
\end{adjustbox}
\label{tab:mutation-score}
\vspace{-0.25cm}
\end{wraptable}

Despite the modest results in this exploratory study, two things are worth noting. First, \tratto can only enhance test suites, not worsen them, as the new oracles generated may reveal actual bugs or increase their mutation score in the best-case scenario. Second, software testing is just one possible application of the oracles generated by \tratto, as they may be useful in other contexts such as program comprehension, requirements specification or runtime verification.

\section{Threats to Validity}
\label{sec:threats}

\noindent
\textbf{Internal validity}.
The datasets used in our experiments may contain wrong instances, since they are based on an automatically generated dataset~\cite{Blasi:Jdoctor:ISSTA:2018}. We mitigate this threat by manually inspecting and fixing or discarding wrong instances. Manually generated or modified instances were inspected by two authors to minimize bias, and conflicts were solved via open discussion. Indeed, all processes involving manual analysis, including the generation of the ground-truth dataset and the analysis of the predictions by \tratto, Jdoctor and GPT4, were performed by two authors. All our datasets, results, and tool implementations are available as open source in our replication package~\cite{supplementarymaterial}.

In answering RQ\textsubscript{1} and RQ\textsubscript{2}, we measure the differences across the techniques (code models and ablated approaches) in terms of the accuracy in generating the next oracle token. This does not take into account equivalent valid oracles that the model can generate. For example, the oracle ``\codeid{param < 0;}'' is equivalent to ``\codeid{(param < 0);}'', thus both tokens `\codeid{(}' and `\codeid{param}' are valid when the partial oracle is empty. However, only `\codeid{param}' is deemed as correct.
The significant difference between the accuracies of the three models in RQ\textsubscript{1} relieves the risk of missing the best LLM for \tratto. As for RQ\textsubscript{2}, we noticed that this phenomenon occurred in both cases (for \tratto and the ablated approaches), partially alleviating this threat to the validity of the results. In RQ\textsubscript{3} and RQ\textsubscript{4} we manually analyzed all oracles generated to fully neutralize this threat and properly compare \tratto against the state-of-the-art approaches, while RQ\textsubscript{5} is not affected by this threat.

In generating the ground-truth dataset, we decided to keep only those methods featuring at least one oracle, otherwise the dataset would be extremely unbalanced, containing 389 positive samples (oracles) and 6,485 negative samples (units for which no oracles can be extracted). Furthermore, this would make the manual analysis extremely costly, due to the myriad of false positives generated by GPT4. We did analyze the results for \tratto and Jdoctor considering also the 6,485 negative samples and found that the recall for both remained in similar levels (49\% for \tratto and 19\% for Jdoctor) while precision decreased (49\% for \tratto and 57\% for Jdoctor). This was expected due to the higher amount of negative samples, making both approaches incur in more false positives.

\vspace{0.2cm}
\noindent
\textbf{External validity}.
Our experiments with Java methods do not prove the generalizability of \tratto to other programming languages. Nevertheless, the approach can be easily adapted to other languages since it merely relies on source code and documentation of inputs and outputs. The symbolic component simply needs a language grammar and context restrictions, while for the neural module an ML model pre-trained on the target programming language is sufficient.

\tratto takes about five times more to analyze a Java class and generate oracles for it, compared to Jdoctor. This is partly because \tratto generates more oracles. Generating oracles token by token causes an overhead, but acceptable for the improvement achieved. Scalability can be addressed as \tratto's neural module is decoupled from the symbolic module, and the communication is handled through a REST API~\cite{richardson13-book,fielding00-phd}. In practice, this means that the neural module could be deployed in a more powerful server, leveraging a bigger code model, to speed up oracle generation.

\section{Related Work}
\label{sec:related}

Our work combines symbolic and neural techniques to generate axiomatic oracles. 
Neuro-symbolic approaches have been studied for some SE problems, like code completion~\cite{Kim:neuro-symbolic-code-prediction:icse:2021, Svyatkovskiy:neuro-symbolic-code-completion:msr:2021, Bibaev:neuro-symbolic:code-completion:sigsoft:2022} or program synthesis~\cite{parisotto2016neuro,cambronero2023flashfill++}. 
To the best of our knowledge, \tratto is the first neuro-symbolic approach tailored to the oracle problem. Next, we discuss both symbolic and neural techniques for oracle generation.

\subsection{Symbolic Approaches}
Symbolic approaches leverage static and dynamic program analysis and natural language processing (NLP) to understand the semantics of software systems and produce semantically relevant oracles.

The seminal approaches Daikon~\cite{Ernst:Daikon:SCP:2007} and Dysy~\cite{Csallner:Dysy:ICSE:2008} on invariant mining execute a program on a collection of inputs against a collection of potential invariants. The accuracy of the invariants inferred with these methods depends on both the quality and completeness of the test cases and the collection of potential invariants provided. Evospex~\cite{Molina:Evospex:ICSE:2021} overcomes the limitations of the previous techniques by exercising the unit under test through its APIs to generate valid pre and post states, without requiring any specification or test. The valid pre and post states undergo mutations that lead to corresponding invalid pre and post states and a genetic algorithm infers valid postconditions guided by the valid/invalid states. GAssert~\cite{Terragni:gassert:ESECFSE:2020} improves automatically inferred assertion oracles, by reducing false positives and negatives with an evolutionary algorithm. All these approaches derive test oracles, by both relying on the execution of the current version of a program and generating regression oracles that cannot detect if the bug is already present in the program.

Text-driven specification mining methods exploit NLP, pattern, semantic, and syntax matching to generate test oracles from code comments and text documentation.
ALICS~\cite{Pandita:inferring:ICSE:2011} mines code contracts in the form of pre- and postconditions from code comments. ALICS uses NLP-based pattern matching with an application domain specific dictionary which hardly generalizes. 
@tComment~\cite{tan:tcomment:ICST:2012} infers null-deference properties of parameters from Javadoc comments, with natural language patterns and heuristics. The patterns are quite narrow and do not generalize to other exception types.

Toradocu~\cite{Goffi:Exceptional:ISSTA:2016}, JDoctor~\cite{Blasi:Jdoctor:ISSTA:2018}, MeMo \cite{Blasi:memo:JSS:2021} and CaMeMa~\cite{Blasi:camema:ASE:2022} use natural language parsing and lexical matching to generate preconditions, normal and exceptional postconditions, metamorphic relations, and temporal properties, from Javadoc comments. 
These approaches can infer oracles when code comments fit the defined patterns, but do not generalize when comments fall outside those. These approaches are not applicable when code comments are unavailable.  We discussed the experimental comparison of \Tratto with Jdoctor, the most complete of this family of approaches, in RQ\textsubscript{3,4,5}.

\subsection{Neural Approaches} 
Recently, there has been an upsurge of works that exploit DL for SE and software testing, with many applications of neural-based transformers and transfer learning to automatically generate semantically relevant assertion oracles and entire unit tests, overcoming the limitations of symbolic techniques even in the absence of precise patterns or entire docstrings.

ATLAS~\cite{Watson:AssertionGenerationNN:ICSE:2020} leverages a recurrent neural network (RNN) to generate assertion oracles from a unit under test and a test prefix. ATLAS relies only on the source code and ignores the documentation. It solely targets normal postconditions and cannot generate oracles for preconditions or exceptional behavior. Subsequently, Tufano et al.~\cite{Tufano:TestCase:2021} proposed AthenaTest, which outperformed ATLAS by replacing the RNN with transformer models pre-trained on natural language and code. 
This was the first noteworthy approach to generate test cases including both test prefixes and oracles, considering both the unit implementation and its context, such as the surrounding class and method signature. AthenaTest does not consider the information provided in the code documentation. 

Dinella et al.~\cite{Dinella:TOGA:ICSE:2022} redesigned the oracle generation problem as a two-step neural ranking procedure. Their approach, TOGA, exploits two pre-trained models fine-tuned on the task of discerning among normal and exceptional behaviors to generate test oracles. The exceptional oracle classifier infers whether a test prefix should throw an exception, from the unit under test, the docstring (if available) and a test prefix.  The assertion oracle ranker generates an assertion oracle conforming to a grammar of candidate assertions, which enforces syntactic and type correctness.
The wide applicability and high flexibility of neural approaches comes with a cost in terms of accuracy. TOGA mitigates the issue by restricting the approach to a subset of plausible and syntactically valid oracles. Hossain et al.~\cite{Hossain:TOG-evaluation:sigsoft:2023} analyzed the current neural-based test oracle generation approaches (including TOGA) and highlighted how the inferred assertions still exhibit high false positive rates that threaten their practical usefulness. Our proposed approach tackles this limitation.

All neural approaches discussed generate concrete assertions as normal and exceptional postconditions. 
\Tratto is the first attempt to successfully generate axiomatic oracles, including preconditions, to invalidate tests when input values are not satisfied, thus mitigating false positives. %

\section{Conclusion}
\label{sec:conclusion}

In this paper, we propose \Tratto, a neuro-symbolic approach to automatically generate axiomatic oracles from commonly available information: documentation and code.
Axiomatic oracles in the form of pre- and postconditions are particularly useful as a complement to automatically generated test suites, as preconditions can rule out invalid inputs, reducing false positives, and postconditions can detect bugs, reducing false negatives. Moreover, axiomatic oracles are applicable to any test case, as they predicate on input and output variables and states.

\tratto entails a novel reformulation of the oracle generation problem, as a token-by-token generation approach. The symbolic module of \tratto effectively restricts the search space of the tokens that may be used to generate an oracle, while the neural component steers the generation process towards semantically relevant oracles. Our ablation studies confirm the contributions of both the symbolic and neural components over solely symbolic- or neural-based approaches. \tratto generates over three times more correct oracles than state-of-the-art symbolic approaches (Jdoctor~\cite{Blasi:Jdoctor:ISSTA:2018}) while incurring in 10 times less false positives than neural approaches (GPT4 complemented with few-shot learning and Chain-of-Thought prompting). Also, \tratto is significantly robust to the quality of documentation, and it can help improve the quality of existing test suites.

Our future research agenda will focus on applying neuro-symbolic techniques to generating concrete assertions and complete test cases.

\section*{Data Availability}
Our replication package includes the source code of the scripts and programs developed, the data generated in the experiments (including datasets) and instructions on how to reuse the material for further research. The artifact can be downloaded from~\cite{supplementarymaterial}.

\begin{acks}
This research was supported in part by NSF grant CNS-2120070, DARPA contract FA8750-20-C-0226, and SNF project A-Test Autonomic Software Testing ({200021\_215487}). 
\end{acks}

\bibliographystyle{ACM-Reference-Format}
\bibliography{bibliography,plume-bib/bibstring-unabbrev,plume-bib/ernst,plume-bib/nlp,plume-bib/pag,plume-bib/testing,plume-bib/crossrefs}


\begin{thebibliography}{60}


\ifx \showCODEN    \undefined \def \showCODEN     #1{\unskip}     \fi
\ifx \showDOI      \undefined \def \showDOI       #1{#1}\fi
\ifx \showISBNx    \undefined \def \showISBNx     #1{\unskip}     \fi
\ifx \showISBNxiii \undefined \def \showISBNxiii  #1{\unskip}     \fi
\ifx \showISSN     \undefined \def \showISSN      #1{\unskip}     \fi
\ifx \showLCCN     \undefined \def \showLCCN      #1{\unskip}     \fi
\ifx \shownote     \undefined \def \shownote      #1{#1}          \fi
\ifx \showarticletitle \undefined \def \showarticletitle #1{#1}   \fi
\ifx \showURL      \undefined \def \showURL       {\relax}        \fi
\providecommand\bibfield[2]{#2}
\providecommand\bibinfo[2]{#2}
\providecommand\natexlab[1]{#1}
\providecommand\showeprint[2][]{arXiv:#2}

\bibitem[jdo(nd)]%
        {jdoctor-dataset}
 \bibinfo{year}{[n.d.]}\natexlab{}.
\newblock \bibinfo{title}{{Dataset of procedure specifications by Blasi et al.}}
\newblock \bibinfo{howpublished}{\url{https://github.com/albertogoffi/toradocu/tree/master/src/test/resources/goal-output}}.
\newblock


\bibitem[Anonymous(2025)]%
        {supplementarymaterial}
\bibfield{author}{\bibinfo{person}{Anonymous}.} \bibinfo{year}{2025}\natexlab{}.
\newblock \bibinfo{title}{{[Replication Package] Tratto: A Neuro-Symbolic Approach to Deriving Axiomatic Test Oracles}}.
\newblock \bibinfo{howpublished}{\url{https://anonymous.4open.science/r/tratto-replication-package-31F6}}.
\newblock


\bibitem[Antoy and Hamlet(2000)]%
        {antoy:testOracles:TSE:2000}
\bibfield{author}{\bibinfo{person}{Sergio Antoy} {and} \bibinfo{person}{Dick Hamlet}.} \bibinfo{year}{2000}\natexlab{}.
\newblock \showarticletitle{Automatically Checking an Implementation against Its Formal Specification}.
\newblock \bibinfo{journal}{\emph{IEEE Transactions on Software Engineering}} \bibinfo{volume}{26}, \bibinfo{number}{1} (\bibinfo{year}{2000}), \bibinfo{pages}{55--69}.
\newblock


\bibitem[Barr et~al\mbox{.}(2015)]%
        {barr:survey:tse:2015}
\bibfield{author}{\bibinfo{person}{Earl~T. Barr}, \bibinfo{person}{Mark Harman}, \bibinfo{person}{Phil McMinn}, \bibinfo{person}{Muzammil Shahbaz}, {and} \bibinfo{person}{Shin Yoo}.} \bibinfo{year}{2015}\natexlab{}.
\newblock \showarticletitle{The Oracle Problem in Software Testing: {A} Survey}.
\newblock \bibinfo{journal}{\emph{IEEE Transactions on Software Engineering}} \bibinfo{volume}{41}, \bibinfo{number}{5} (\bibinfo{year}{2015}), \bibinfo{pages}{507--525}.
\newblock


\bibitem[Bibaev et~al\mbox{.}(2022)]%
        {Bibaev:neuro-symbolic:code-completion:sigsoft:2022}
\bibfield{author}{\bibinfo{person}{Vitaliy Bibaev}, \bibinfo{person}{Alexey Kalina}, \bibinfo{person}{Vadim Lomshakov}, \bibinfo{person}{Yaroslav Golubev}, \bibinfo{person}{Alexander Bezzubov}, \bibinfo{person}{Nikita Povarov}, {and} \bibinfo{person}{Timofey Bryksin}.} \bibinfo{year}{2022}\natexlab{}.
\newblock \showarticletitle{All you need is logs: improving code completion by learning from anonymous {IDE} usage logs}. In \bibinfo{booktitle}{\emph{Proceedings of the 30th {ACM} Joint European Software Engineering Conference and Symposium on the Foundations of Software Engineering, {ESEC/FSE} 2022, Singapore, Singapore, November 14-18, 2022}}, \bibfield{editor}{\bibinfo{person}{Abhik Roychoudhury}, \bibinfo{person}{Cristian Cadar}, {and} \bibinfo{person}{Miryung Kim}} (Eds.). \bibinfo{publisher}{{ACM}}, \bibinfo{pages}{1269--1279}.
\newblock
\urldef\tempurl%
\url{https://doi.org/10.1145/3540250.3558968}
\showDOI{\tempurl}


\bibitem[Blasi et~al\mbox{.}(2018)]%
        {Blasi:Jdoctor:ISSTA:2018}
\bibfield{author}{\bibinfo{person}{Arianna Blasi}, \bibinfo{person}{Alberto Goffi}, \bibinfo{person}{Konstantin Kuznetsov}, \bibinfo{person}{Alessandra Gorla}, \bibinfo{person}{Michael~D. Ernst}, \bibinfo{person}{Mauro Pezz{\`e}}, {and} \bibinfo{person}{Sergio Delgado~Castellanos}.} \bibinfo{year}{2018}\natexlab{}.
\newblock \showarticletitle{Translating Code Comments to Procedure Specifications}. In \bibinfo{booktitle}{\emph{Proceedings of the International Symposium on Software Testing and Analysis}} \emph{(\bibinfo{series}{ISSTA~'18})}. \bibinfo{publisher}{ACM}.
\newblock


\bibitem[Blasi et~al\mbox{.}(2022)]%
        {Blasi:camema:ASE:2022}
\bibfield{author}{\bibinfo{person}{Arianna Blasi}, \bibinfo{person}{Alessandra Gorla}, \bibinfo{person}{Michael~D. Ernst}, {and} \bibinfo{person}{Mauro Pezz{\`{e}}}.} \bibinfo{year}{2022}\natexlab{}.
\newblock \showarticletitle{Call Me Maybe: Using {NLP} to Automatically Generate Unit Test Cases Respecting Temporal Constraints}. In \bibinfo{booktitle}{\emph{37th {IEEE/ACM} International Conference on Automated Software Engineering, {ASE} 2022, Rochester, MI, USA, October 10-14, 2022}}. \bibinfo{publisher}{{ACM}}, \bibinfo{pages}{19:1--19:11}.
\newblock
\urldef\tempurl%
\url{https://doi.org/10.1145/3551349.3556961}
\showDOI{\tempurl}


\bibitem[Blasi et~al\mbox{.}(2021)]%
        {Blasi:memo:JSS:2021}
\bibfield{author}{\bibinfo{person}{Arianna Blasi}, \bibinfo{person}{Alessandra Gorla}, \bibinfo{person}{Michael~D Ernst}, \bibinfo{person}{Mauro Pezze}, {and} \bibinfo{person}{Antonio Carzaniga}.} \bibinfo{year}{2021}\natexlab{}.
\newblock \showarticletitle{MeMo: Automatically identifying metamorphic relations in Javadoc comments for test automation}.
\newblock \bibinfo{journal}{\emph{Journal of Systems and Software}}  \bibinfo{volume}{181} (\bibinfo{year}{2021}), \bibinfo{pages}{111041}.
\newblock


\bibitem[Cambronero et~al\mbox{.}(2023)]%
        {cambronero2023flashfill++}
\bibfield{author}{\bibinfo{person}{Jos{\'e} Cambronero}, \bibinfo{person}{Sumit Gulwani}, \bibinfo{person}{Vu Le}, \bibinfo{person}{Daniel Perelman}, \bibinfo{person}{Arjun Radhakrishna}, \bibinfo{person}{Clint Simon}, {and} \bibinfo{person}{Ashish Tiwari}.} \bibinfo{year}{2023}\natexlab{}.
\newblock \showarticletitle{{FlashFill++: Scaling programming by example by cutting to the chase}}.
\newblock \bibinfo{journal}{\emph{Proceedings of the ACM on Programming Languages}} \bibinfo{volume}{7}, \bibinfo{number}{POPL} (\bibinfo{year}{2023}), \bibinfo{pages}{952--981}.
\newblock


\bibitem[Chen et~al\mbox{.}(1998)]%
        {ChenCY1998}
\bibfield{author}{\bibinfo{person}{T.~Y. Chen}, \bibinfo{person}{S.~C. Cheung}, {and} \bibinfo{person}{S.~M. Yiu}.} \bibinfo{year}{1998}\natexlab{}.
\newblock \bibinfo{booktitle}{\emph{Metamorphic testing: A new approach for generating next test cases}}.
\newblock \bibinfo{type}{{T}echnical {R}eport} HKUST-CS98-01. \bibinfo{institution}{HKUST Department of Computer Science}, \bibinfo{address}{Hong Kong}.
\newblock


\bibitem[Cheng and Atlee(2007)]%
        {cheng2007research}
\bibfield{author}{\bibinfo{person}{Betty~HC Cheng} {and} \bibinfo{person}{Joanne~M Atlee}.} \bibinfo{year}{2007}\natexlab{}.
\newblock \showarticletitle{Research directions in requirements engineering}.
\newblock \bibinfo{journal}{\emph{Future of software engineering (FOSE'07)}} (\bibinfo{year}{2007}), \bibinfo{pages}{285--303}.
\newblock


\bibitem[Cheon and Leavens(2002a)]%
        {cheon:jmlJunit:ECOOP:2002}
\bibfield{author}{\bibinfo{person}{Yoonsik Cheon} {and} \bibinfo{person}{Gary~T. Leavens}.} \bibinfo{year}{2002}\natexlab{a}.
\newblock \showarticletitle{A Simple and Practical Approach to Unit Testing: The {JML} and {JUnit} Way}. In \bibinfo{booktitle}{\emph{Proceedings of the European Conference on Object-Oriented Programming}} \emph{(\bibinfo{series}{ECOOP '02})}. \bibinfo{pages}{231--255}.
\newblock


\bibitem[Cheon and Leavens(2002b)]%
        {CheonL2002}
\bibfield{author}{\bibinfo{person}{Yoonsik Cheon} {and} \bibinfo{person}{Gary~T. Leavens}.} \bibinfo{year}{2002}\natexlab{b}.
\newblock \showarticletitle{A simple and practical approach to unit testing: The {JML} and {JUnit} way}. In \bibinfo{booktitle}{\emph{ECOOP 2002 --- Object-Oriented Programming, 16th European Conference}}. \bibinfo{address}{M{\'a}laga, Spain}, \bibinfo{pages}{231--255}.
\newblock


\bibitem[Claessen and Hughes(2000)]%
        {ClaessenH2000}
\bibfield{author}{\bibinfo{person}{Koen Claessen} {and} \bibinfo{person}{John Hughes}.} \bibinfo{year}{2000}\natexlab{}.
\newblock \showarticletitle{{QuickCheck}: A lightweight tool for random testing of {Haskell} programs}. In \bibinfo{booktitle}{\emph{ICFP 2000: Proceedings of the fifth ACM SIGPLAN International Conference on Functional Programming}}. \bibinfo{address}{Montreal, Canada}, \bibinfo{pages}{268--279}.
\newblock


\bibitem[Coles et~al\mbox{.}(2016)]%
        {coles2016pit}
\bibfield{author}{\bibinfo{person}{Henry Coles}, \bibinfo{person}{Thomas Laurent}, \bibinfo{person}{Christopher Henard}, \bibinfo{person}{Mike Papadakis}, {and} \bibinfo{person}{Anthony Ventresque}.} \bibinfo{year}{2016}\natexlab{}.
\newblock \showarticletitle{{PIT: a practical mutation testing tool for Java}}. In \bibinfo{booktitle}{\emph{Proceedings of the 25th international symposium on software testing and analysis}}. \bibinfo{pages}{449--452}.
\newblock


\bibitem[Cornelissen et~al\mbox{.}(2009)]%
        {cornelissen2009systematic}
\bibfield{author}{\bibinfo{person}{Bas Cornelissen}, \bibinfo{person}{Andy Zaidman}, \bibinfo{person}{Arie Van~Deursen}, \bibinfo{person}{Leon Moonen}, {and} \bibinfo{person}{Rainer Koschke}.} \bibinfo{year}{2009}\natexlab{}.
\newblock \showarticletitle{A systematic survey of program comprehension through dynamic analysis}.
\newblock \bibinfo{journal}{\emph{IEEE Transactions on Software Engineering}} \bibinfo{volume}{35}, \bibinfo{number}{5} (\bibinfo{year}{2009}), \bibinfo{pages}{684--702}.
\newblock


\bibitem[Csallner et~al\mbox{.}(2008)]%
        {Csallner:Dysy:ICSE:2008}
\bibfield{author}{\bibinfo{person}{Christoph Csallner}, \bibinfo{person}{Nikolai Tillmann}, {and} \bibinfo{person}{Yannis Smaragdakis}.} \bibinfo{year}{2008}\natexlab{}.
\newblock \showarticletitle{{DySy}: Dynamic Symbolic Execution for Invariant Inference}. In \bibinfo{booktitle}{\emph{Proceedings of the International Conference on Software Engineering}} \emph{(\bibinfo{series}{ICSE '08})}. \bibinfo{publisher}{ACM}, \bibinfo{pages}{281--290}.
\newblock


\bibitem[Day and Gannon(1985)]%
        {day:oraclesSpecifications:SOFTAIR:1985}
\bibfield{author}{\bibinfo{person}{J.~D. Day} {and} \bibinfo{person}{J.~D. Gannon}.} \bibinfo{year}{1985}\natexlab{}.
\newblock \showarticletitle{A Test Oracle Based on Formal Specifications}. In \bibinfo{booktitle}{\emph{Proceedings of the Conference on Software Development Tools, Techniques, and Alternatives}} \emph{(\bibinfo{series}{SOFTAIR '85})}. \bibinfo{pages}{126--130}.
\newblock


\bibitem[Dinella et~al\mbox{.}(2022)]%
        {Dinella:TOGA:ICSE:2022}
\bibfield{author}{\bibinfo{person}{Elizabeth Dinella}, \bibinfo{person}{Gabriel Ryan}, \bibinfo{person}{Todd Mytkowicz}, {and} \bibinfo{person}{Shuvendu Lahiri}.} \bibinfo{year}{2022}\natexlab{}.
\newblock \showarticletitle{TOGA: A Neural Method for Test Oracle Generation}. In \bibinfo{booktitle}{\emph{ICSE 2022}}. ACM.
\newblock
\urldef\tempurl%
\url{https://www.microsoft.com/en-us/research/publication/toga-a-neural-method-for-test-oracle-generation/}
\showURL{%
\tempurl}


\bibitem[Ernst et~al\mbox{.}(2007)]%
        {Ernst:Daikon:SCP:2007}
\bibfield{author}{\bibinfo{person}{Michael~D. Ernst}, \bibinfo{person}{Jeff~H. Perkins}, \bibinfo{person}{Philip~J. Guo}, \bibinfo{person}{Stephen McCamant}, \bibinfo{person}{Carlos Pacheco}, \bibinfo{person}{Matthew~S. Tschantz}, {and} \bibinfo{person}{Chen Xiao}.} \bibinfo{year}{2007}\natexlab{}.
\newblock \showarticletitle{The {Daikon} system for dynamic detection of likely invariants}.
\newblock \bibinfo{journal}{\emph{Science of Computer Programming}} \bibinfo{volume}{69}, \bibinfo{number}{1--3} (\bibinfo{year}{2007}), \bibinfo{pages}{35--45}.
\newblock


\bibitem[Fielding(2000)]%
        {fielding00-phd}
\bibfield{author}{\bibinfo{person}{Roy~Thomas Fielding}.} \bibinfo{year}{2000}\natexlab{}.
\newblock \emph{\bibinfo{title}{{Architectural Styles and the Design of Network-based Software Architectures}}}.
\newblock \bibinfo{thesistype}{Ph.\,D. Dissertation}.
\newblock
\showISBNx{0-599-87118-0}


\bibitem[Fink and Bishop(1997)]%
        {FinkB1997}
\bibfield{author}{\bibinfo{person}{George Fink} {and} \bibinfo{person}{Matt Bishop}.} \bibinfo{year}{1997}\natexlab{}.
\newblock \showarticletitle{Property-based testing: A new approach to testing for assurance}.
\newblock \bibinfo{journal}{\emph{ACM SIGSOFT Software Engineering Notes}} \bibinfo{volume}{22}, \bibinfo{number}{4} (\bibinfo{date}{July} \bibinfo{year}{1997}), \bibinfo{pages}{74--80}.
\newblock


\bibitem[Foundation(2023a)]%
        {apache_commons_collections}
\bibfield{author}{\bibinfo{person}{Apache~Software Foundation}.} \bibinfo{year}{2023}\natexlab{a}.
\newblock \bibinfo{title}{Apache Commons Collections: The Apache Commons Collections Library}.
\newblock \bibinfo{howpublished}{\url{https://commons.apache.org/proper/commons-collections/}}.
\newblock
\newblock
\shownote{Accessed: 2024-06-07}.


\bibitem[Foundation(2023b)]%
        {apache_commons_math}
\bibfield{author}{\bibinfo{person}{Apache~Software Foundation}.} \bibinfo{year}{2023}\natexlab{b}.
\newblock \bibinfo{title}{Apache Commons Math: The Apache Commons Mathematics Library}.
\newblock \bibinfo{howpublished}{\url{https://commons.apache.org/proper/commons-math/}}.
\newblock
\newblock
\shownote{Accessed: 2024-06-07}.


\bibitem[Fraser and Arcuri(2011)]%
        {Fraser:EvoSuite:ESECFSE:2011}
\bibfield{author}{\bibinfo{person}{Gordon Fraser} {and} \bibinfo{person}{Andrea Arcuri}.} \bibinfo{year}{2011}\natexlab{}.
\newblock \showarticletitle{EvoSuite: Automatic Test Suite Generation for Object-Oriented Software}. In \bibinfo{booktitle}{\emph{Proceedings of the European Software Engineering Conference held jointly with the ACM SIGSOFT International Symposium on Foundations of Software Engineering}} \emph{(\bibinfo{series}{ESEC/FSE '11})}. \bibinfo{publisher}{ACM}, \bibinfo{pages}{416--419}.
\newblock


\bibitem[Fraser and Arcuri(2013)]%
        {Fraser:Evosuite:TSE:2013}
\bibfield{author}{\bibinfo{person}{Gordon Fraser} {and} \bibinfo{person}{Andrea Arcuri}.} \bibinfo{year}{2013}\natexlab{}.
\newblock \showarticletitle{Whole Test Suite Generation}.
\newblock \bibinfo{journal}{\emph{IEEE Transactions on Software Engineering}} \bibinfo{volume}{39}, \bibinfo{number}{2} (\bibinfo{year}{2013}), \bibinfo{pages}{276--291}.
\newblock


\bibitem[Goffi et~al\mbox{.}(2016)]%
        {Goffi:Exceptional:ISSTA:2016}
\bibfield{author}{\bibinfo{person}{Alberto Goffi}, \bibinfo{person}{Alessandra Gorla}, \bibinfo{person}{Michael~D. Ernst}, {and} \bibinfo{person}{Mauro Pezz{\`e}}.} \bibinfo{year}{2016}\natexlab{}.
\newblock \showarticletitle{Automatic Generation of Oracles for Exceptional Behaviors}. In \bibinfo{booktitle}{\emph{Proceedings of the International Symposium on Software Testing and Analysis}} \emph{(\bibinfo{series}{ISSTA~'16})}. \bibinfo{publisher}{ACM}, \bibinfo{pages}{213--224}.
\newblock


\bibitem[Hossain et~al\mbox{.}(2023)]%
        {Hossain:TOG-evaluation:sigsoft:2023}
\bibfield{author}{\bibinfo{person}{Soneya~Binta Hossain}, \bibinfo{person}{Antonio Filieri}, \bibinfo{person}{Matthew~B. Dwyer}, \bibinfo{person}{Sebastian~G. Elbaum}, {and} \bibinfo{person}{Willem Visser}.} \bibinfo{year}{2023}\natexlab{}.
\newblock \showarticletitle{Neural-Based Test Oracle Generation: {A} Large-Scale Evaluation and Lessons Learned}. In \bibinfo{booktitle}{\emph{Proceedings of the 31st {ACM} Joint European Software Engineering Conference and Symposium on the Foundations of Software Engineering, {ESEC/FSE} 2023, San Francisco, CA, USA, December 3-9, 2023}}, \bibfield{editor}{\bibinfo{person}{Satish Chandra}, \bibinfo{person}{Kelly Blincoe}, {and} \bibinfo{person}{Paolo Tonella}} (Eds.). \bibinfo{publisher}{{ACM}}, \bibinfo{pages}{120--132}.
\newblock
\urldef\tempurl%
\url{https://doi.org/10.1145/3611643.3616265}
\showDOI{\tempurl}


\bibitem[Jia and Harman(2011)]%
        {Jia:Analysis:TSE:2011}
\bibfield{author}{\bibinfo{person}{Yue Jia} {and} \bibinfo{person}{Mark Harman}.} \bibinfo{year}{2011}\natexlab{}.
\newblock \showarticletitle{An Analysis and Survey of the Development of Mutation Testing}.
\newblock \bibinfo{journal}{\emph{IEEE Transactions on Software Engineering}} \bibinfo{volume}{37}, \bibinfo{number}{5} (\bibinfo{date}{September} \bibinfo{year}{2011}), \bibinfo{pages}{649--678}.
\newblock


\bibitem[Just et~al\mbox{.}(2014)]%
        {Just:Defects4j:ISSTA:2014}
\bibfield{author}{\bibinfo{person}{Ren{\'e} Just}, \bibinfo{person}{Darioush Jalali}, {and} \bibinfo{person}{Michael~D. Ernst}.} \bibinfo{year}{2014}\natexlab{}.
\newblock \showarticletitle{{Defects4J}: A Database of Existing Faults to Enable Controlled Testing Studies for {J}ava Programs}. In \bibinfo{booktitle}{\emph{Proceedings of the International Symposium on Software Testing and Analysis}} \emph{(\bibinfo{series}{ISSTA '14})}. \bibinfo{publisher}{ACM}, \bibinfo{pages}{437--440}.
\newblock


\bibitem[Kanstren(2009)]%
        {kanstren2009program}
\bibfield{author}{\bibinfo{person}{Teemu Kanstren}.} \bibinfo{year}{2009}\natexlab{}.
\newblock \showarticletitle{Program comprehension for user-assisted test oracle generation}. In \bibinfo{booktitle}{\emph{2009 Fourth International Conference on Software Engineering Advances}}. IEEE, \bibinfo{pages}{118--127}.
\newblock


\bibitem[Kim et~al\mbox{.}(2021)]%
        {Kim:neuro-symbolic-code-prediction:icse:2021}
\bibfield{author}{\bibinfo{person}{Seohyun Kim}, \bibinfo{person}{Jinman Zhao}, \bibinfo{person}{Yuchi Tian}, {and} \bibinfo{person}{Satish Chandra}.} \bibinfo{year}{2021}\natexlab{}.
\newblock \showarticletitle{Code Prediction by Feeding Trees to Transformers}. In \bibinfo{booktitle}{\emph{43rd {IEEE/ACM} International Conference on Software Engineering, {ICSE} 2021, Madrid, Spain, 22-30 May 2021}}. \bibinfo{publisher}{{IEEE}}, \bibinfo{pages}{150--162}.
\newblock
\urldef\tempurl%
\url{https://doi.org/10.1109/ICSE43902.2021.00026}
\showDOI{\tempurl}


\bibitem[Leucker and Schallhart(2009)]%
        {leucker2009brief}
\bibfield{author}{\bibinfo{person}{Martin Leucker} {and} \bibinfo{person}{Christian Schallhart}.} \bibinfo{year}{2009}\natexlab{}.
\newblock \showarticletitle{A brief account of runtime verification}.
\newblock \bibinfo{journal}{\emph{The journal of logic and algebraic programming}} \bibinfo{volume}{78}, \bibinfo{number}{5} (\bibinfo{year}{2009}), \bibinfo{pages}{293--303}.
\newblock


\bibitem[Lozhkov et~al\mbox{.}(2024)]%
        {lozhkov2024starcoder2stackv2}
\bibfield{author}{\bibinfo{person}{Anton Lozhkov}, \bibinfo{person}{Raymond Li}, \bibinfo{person}{Loubna~Ben Allal}, \bibinfo{person}{Federico Cassano}, \bibinfo{person}{Joel Lamy-Poirier}, \bibinfo{person}{Nouamane Tazi}, \bibinfo{person}{Ao Tang}, \bibinfo{person}{Dmytro Pykhtar}, \bibinfo{person}{Jiawei Liu}, \bibinfo{person}{Yuxiang Wei}, \bibinfo{person}{Tianyang Liu}, \bibinfo{person}{Max Tian}, \bibinfo{person}{Denis Kocetkov}, \bibinfo{person}{Arthur Zucker}, \bibinfo{person}{Younes Belkada}, \bibinfo{person}{Zijian Wang}, \bibinfo{person}{Qian Liu}, \bibinfo{person}{Dmitry Abulkhanov}, \bibinfo{person}{Indraneil Paul}, \bibinfo{person}{Zhuang Li}, \bibinfo{person}{Wen-Ding Li}, \bibinfo{person}{Megan Risdal}, \bibinfo{person}{Jia Li}, \bibinfo{person}{Jian Zhu}, \bibinfo{person}{Terry~Yue Zhuo}, \bibinfo{person}{Evgenii Zheltonozhskii}, \bibinfo{person}{Nii Osae~Osae Dade}, \bibinfo{person}{Wenhao Yu}, \bibinfo{person}{Lucas Krauß}, \bibinfo{person}{Naman Jain}, \bibinfo{person}{Yixuan Su}, \bibinfo{person}{Xuanli He}, \bibinfo{person}{Manan Dey}, \bibinfo{person}{Edoardo Abati}, \bibinfo{person}{Yekun Chai}, \bibinfo{person}{Niklas Muennighoff}, \bibinfo{person}{Xiangru Tang}, \bibinfo{person}{Muhtasham Oblokulov}, \bibinfo{person}{Christopher Akiki}, \bibinfo{person}{Marc Marone}, \bibinfo{person}{Chenghao Mou}, \bibinfo{person}{Mayank Mishra}, \bibinfo{person}{Alex Gu}, \bibinfo{person}{Binyuan Hui}, \bibinfo{person}{Tri Dao}, \bibinfo{person}{Armel Zebaze}, \bibinfo{person}{Olivier Dehaene}, \bibinfo{person}{Nicolas Patry}, \bibinfo{person}{Canwen Xu}, \bibinfo{person}{Julian McAuley}, \bibinfo{person}{Han Hu}, \bibinfo{person}{Torsten Scholak}, \bibinfo{person}{Sebastien Paquet}, \bibinfo{person}{Jennifer Robinson}, \bibinfo{person}{Carolyn~Jane Anderson}, \bibinfo{person}{Nicolas Chapados}, \bibinfo{person}{Mostofa Patwary}, \bibinfo{person}{Nima Tajbakhsh}, \bibinfo{person}{Yacine Jernite}, \bibinfo{person}{Carlos~Muñoz Ferrandis}, \bibinfo{person}{Lingming Zhang}, \bibinfo{person}{Sean Hughes}, \bibinfo{person}{Thomas Wolf}, \bibinfo{person}{Arjun Guha}, \bibinfo{person}{Leandro von Werra}, {and} \bibinfo{person}{Harm de Vries}.} \bibinfo{year}{2024}\natexlab{}.
\newblock \bibinfo{title}{StarCoder 2 and The Stack v2: The Next Generation}.
\newblock
\newblock
\showeprint[arxiv]{2402.19173}~[cs.SE]
\urldef\tempurl%
\url{https://arxiv.org/abs/2402.19173}
\showURL{%
\tempurl}


\bibitem[Mastropaolo et~al\mbox{.}(2023)]%
        {mastropaolo:learning:press:2023}
\bibfield{author}{\bibinfo{person}{Antonio Mastropaolo}, \bibinfo{person}{Nathan Cooper}, \bibinfo{person}{David~Nader Palacio}, \bibinfo{person}{Simone Scalabrino}, \bibinfo{person}{Denys Poshyvanyk}, \bibinfo{person}{Rocco Oliveto}, {and} \bibinfo{person}{Gabriele Bavota}.} \bibinfo{year}{2023}\natexlab{}.
\newblock \showarticletitle{Using Transfer Learning for Code-Related Tasks}.
\newblock \bibinfo{journal}{\emph{IEEE Trans. Softw. Eng.}} \bibinfo{volume}{49}, \bibinfo{number}{4} (\bibinfo{date}{apr} \bibinfo{year}{2023}), \bibinfo{pages}{1580–1598}.
\newblock
\showISSN{0098-5589}
\urldef\tempurl%
\url{https://doi.org/10.1109/TSE.2022.3183297}
\showDOI{\tempurl}


\bibitem[McNemar(1947)]%
        {mcnemar}
\bibfield{author}{\bibinfo{person}{Quinn McNemar}.} \bibinfo{year}{1947}\natexlab{}.
\newblock \showarticletitle{Note on the sampling error of the difference between correlated proportions or percentages}.
\newblock \bibinfo{journal}{\emph{Psychometrika}} \bibinfo{volume}{12}, \bibinfo{number}{2} (\bibinfo{year}{1947}), \bibinfo{pages}{153--157}.
\newblock


\bibitem[Molina et~al\mbox{.}(2021)]%
        {Molina:Evospex:ICSE:2021}
\bibfield{author}{\bibinfo{person}{Facundo Molina}, \bibinfo{person}{Pablo Ponzio}, \bibinfo{person}{Nazareno Aguirre}, {and} \bibinfo{person}{Marcelo Frias}.} \bibinfo{year}{2021}\natexlab{}.
\newblock \showarticletitle{EvoSpex: An evolutionary algorithm for learning postconditions}. In \bibinfo{booktitle}{\emph{2021 IEEE/ACM 43rd International Conference on Software Engineering (ICSE)}}. IEEE, \bibinfo{pages}{1223--1235}.
\newblock


\bibitem[OpenAI(2023a)]%
        {openai2023chatgpt}
\bibfield{author}{\bibinfo{person}{OpenAI}.} \bibinfo{year}{2023}\natexlab{a}.
\newblock \bibinfo{title}{ChatGPT: An AI Language Model}.
\newblock \bibinfo{howpublished}{\url{https://www.openai.com/chatgpt}}.
\newblock
\newblock
\shownote{Accessed: 2024-06-07}.


\bibitem[OpenAI(2023b)]%
        {OpenAI:GPT4-o:corr:2024}
\bibfield{author}{\bibinfo{person}{OpenAI}.} \bibinfo{year}{2023}\natexlab{b}.
\newblock \showarticletitle{{GPT-4} Technical Report}.
\newblock \bibinfo{journal}{\emph{CoRR}}  \bibinfo{volume}{abs/2303.08774} (\bibinfo{year}{2023}).
\newblock
\urldef\tempurl%
\url{https://doi.org/10.48550/ARXIV.2303.08774}
\showDOI{\tempurl}
\showeprint[arXiv]{2303.08774}


\bibitem[OpenAI(2024)]%
        {gpt4}
\bibfield{author}{\bibinfo{person}{OpenAI}.} \bibinfo{year}{2024}\natexlab{}.
\newblock \bibinfo{title}{GPT-4 | OpenAI}.
\newblock \bibinfo{howpublished}{\url{https://openai.com/index/gpt-4/}}.
\newblock
\newblock
\shownote{Accessed: 2024-10-07}.


\bibitem[Pacheco et~al\mbox{.}(2007)]%
        {Pacheco:Randoop:ICSE:2007}
\bibfield{author}{\bibinfo{person}{Carlos Pacheco}, \bibinfo{person}{Shuvendu~K. Lahiri}, \bibinfo{person}{Michael~D. Ernst}, {and} \bibinfo{person}{Thomas Ball}.} \bibinfo{year}{2007}\natexlab{}.
\newblock \showarticletitle{Feedback-Directed Random Test Generation}. In \bibinfo{booktitle}{\emph{Proceedings of the International Conference on Software Engineering}} \emph{(\bibinfo{series}{ICSE '07})}. \bibinfo{publisher}{ACM}, \bibinfo{pages}{75--84}.
\newblock


\bibitem[Pandita et~al\mbox{.}(2012)]%
        {Pandita:inferring:ICSE:2011}
\bibfield{author}{\bibinfo{person}{Rahul Pandita}, \bibinfo{person}{Xusheng Xiao}, \bibinfo{person}{Hao Zhong}, \bibinfo{person}{Tao Xie}, \bibinfo{person}{Stephen Oney}, {and} \bibinfo{person}{Amit Paradkar}.} \bibinfo{year}{2012}\natexlab{}.
\newblock \showarticletitle{Inferring method specifications from natural language API descriptions}. In \bibinfo{booktitle}{\emph{2012 34th international conference on software engineering (ICSE)}}. IEEE, \bibinfo{pages}{815--825}.
\newblock


\bibitem[Parisotto et~al\mbox{.}(2016)]%
        {parisotto2016neuro}
\bibfield{author}{\bibinfo{person}{Emilio Parisotto}, \bibinfo{person}{Abdel-rahman Mohamed}, \bibinfo{person}{Rishabh Singh}, \bibinfo{person}{Lihong Li}, \bibinfo{person}{Dengyong Zhou}, {and} \bibinfo{person}{Pushmeet Kohli}.} \bibinfo{year}{2016}\natexlab{}.
\newblock \showarticletitle{{Neuro-symbolic program synthesis}}. In \bibinfo{booktitle}{\emph{International Conference on Learning Representations}}.
\newblock


\bibitem[Peters and Parnas(1998)]%
        {Peters:OraclesFromDocumentation:TSE:1998}
\bibfield{author}{\bibinfo{person}{Dennis~K. Peters} {and} \bibinfo{person}{David~Lorge Parnas}.} \bibinfo{year}{1998}\natexlab{}.
\newblock \showarticletitle{Using test oracles generated from program documentation}.
\newblock \bibinfo{journal}{\emph{IEEE Transactions on Software Engineering}} \bibinfo{volume}{24}, \bibinfo{number}{3} (\bibinfo{date}{3} \bibinfo{year}{1998}), \bibinfo{pages}{161--173}.
\newblock


\bibitem[Pezz{\`{e}} and Zhang(2015)]%
        {Pezze:Oracles:Advances:2015}
\bibfield{author}{\bibinfo{person}{Mauro Pezz{\`{e}}} {and} \bibinfo{person}{Cheng Zhang}.} \bibinfo{year}{2015}\natexlab{}.
\newblock \showarticletitle{Automated Test Oracles: {A} Survey}.
\newblock In \bibinfo{booktitle}{\emph{Advances in Computers}}. Vol.~\bibinfo{volume}{95}. \bibinfo{publisher}{Elsevier}, \bibinfo{pages}{1--48}.
\newblock


\bibitem[Richardson et~al\mbox{.}(2013)]%
        {richardson13-book}
\bibfield{author}{\bibinfo{person}{Leonard Richardson}, \bibinfo{person}{Mike Amundsen}, {and} \bibinfo{person}{Sam Ruby}.} \bibinfo{year}{2013}\natexlab{}.
\newblock \bibinfo{booktitle}{\emph{{RESTful Web APIs}}}.
\newblock \bibinfo{publisher}{O'Reilly Media, Inc.}
\newblock
\showISBNx{1449358063, 9781449358068}


\bibitem[Rozière et~al\mbox{.}(2024)]%
        {rozière2024codellamaopenfoundation}
\bibfield{author}{\bibinfo{person}{Baptiste Rozière}, \bibinfo{person}{Jonas Gehring}, \bibinfo{person}{Fabian Gloeckle}, \bibinfo{person}{Sten Sootla}, \bibinfo{person}{Itai Gat}, \bibinfo{person}{Xiaoqing~Ellen Tan}, \bibinfo{person}{Yossi Adi}, \bibinfo{person}{Jingyu Liu}, \bibinfo{person}{Romain Sauvestre}, \bibinfo{person}{Tal Remez}, \bibinfo{person}{Jérémy Rapin}, \bibinfo{person}{Artyom Kozhevnikov}, \bibinfo{person}{Ivan Evtimov}, \bibinfo{person}{Joanna Bitton}, \bibinfo{person}{Manish Bhatt}, \bibinfo{person}{Cristian~Canton Ferrer}, \bibinfo{person}{Aaron Grattafiori}, \bibinfo{person}{Wenhan Xiong}, \bibinfo{person}{Alexandre Défossez}, \bibinfo{person}{Jade Copet}, \bibinfo{person}{Faisal Azhar}, \bibinfo{person}{Hugo Touvron}, \bibinfo{person}{Louis Martin}, \bibinfo{person}{Nicolas Usunier}, \bibinfo{person}{Thomas Scialom}, {and} \bibinfo{person}{Gabriel Synnaeve}.} \bibinfo{year}{2024}\natexlab{}.
\newblock \bibinfo{title}{Code Llama: Open Foundation Models for Code}.
\newblock
\newblock
\showeprint[arxiv]{2308.12950}~[cs.CL]
\urldef\tempurl%
\url{https://arxiv.org/abs/2308.12950}
\showURL{%
\tempurl}


\bibitem[Saff(2007)]%
        {Saff2007:OOPSLA2007demo}
\bibfield{author}{\bibinfo{person}{David Saff}.} \bibinfo{year}{2007}\natexlab{}.
\newblock \showarticletitle{Theory-infected: Or how {I} learned to stop worrying and love universal quantification}. In \bibinfo{booktitle}{\emph{OOPSLA Companion: Object-Oriented Programming Systems, Languages, and Applications}}. \bibinfo{address}{Montreal, Canada}, \bibinfo{pages}{846--847}.
\newblock


\bibitem[Svyatkovskiy et~al\mbox{.}(2021)]%
        {Svyatkovskiy:neuro-symbolic-code-completion:msr:2021}
\bibfield{author}{\bibinfo{person}{Alexey Svyatkovskiy}, \bibinfo{person}{Sebastian Lee}, \bibinfo{person}{Anna Hadjitofi}, \bibinfo{person}{Maik Riechert}, \bibinfo{person}{Juliana~Vicente Franco}, {and} \bibinfo{person}{Miltiadis Allamanis}.} \bibinfo{year}{2021}\natexlab{}.
\newblock \showarticletitle{Fast and Memory-Efficient Neural Code Completion}. In \bibinfo{booktitle}{\emph{18th {IEEE/ACM} International Conference on Mining Software Repositories, {MSR} 2021, Madrid, Spain, May 17-19, 2021}}. \bibinfo{publisher}{{IEEE}}, \bibinfo{pages}{329--340}.
\newblock
\urldef\tempurl%
\url{https://doi.org/10.1109/MSR52588.2021.00045}
\showDOI{\tempurl}


\bibitem[Tan et~al\mbox{.}(2012)]%
        {tan:tcomment:ICST:2012}
\bibfield{author}{\bibinfo{person}{Shin~Hwei Tan}, \bibinfo{person}{Darko Marinov}, \bibinfo{person}{Lin Tan}, {and} \bibinfo{person}{Gary~T. Leavens}.} \bibinfo{year}{2012}\natexlab{}.
\newblock \showarticletitle{{@tComment}: Testing {J}avadoc Comments to Detect Comment-Code Inconsistencies}. In \bibinfo{booktitle}{\emph{Proceedings of the International Conference on Software Testing, Verification and Validation}} \emph{(\bibinfo{series}{ICST '12})}. \bibinfo{publisher}{IEEE Computer Society}, \bibinfo{pages}{260--269}.
\newblock


\bibitem[Team et~al\mbox{.}(2024)]%
        {codegemmateam2024codegemmaopencodemodels}
\bibfield{author}{\bibinfo{person}{CodeGemma Team}, \bibinfo{person}{Heri Zhao}, \bibinfo{person}{Jeffrey Hui}, \bibinfo{person}{Joshua Howland}, \bibinfo{person}{Nam Nguyen}, \bibinfo{person}{Siqi Zuo}, \bibinfo{person}{Andrea Hu}, \bibinfo{person}{Christopher~A. Choquette-Choo}, \bibinfo{person}{Jingyue Shen}, \bibinfo{person}{Joe Kelley}, \bibinfo{person}{Kshitij Bansal}, \bibinfo{person}{Luke Vilnis}, \bibinfo{person}{Mateo Wirth}, \bibinfo{person}{Paul Michel}, \bibinfo{person}{Peter Choy}, \bibinfo{person}{Pratik Joshi}, \bibinfo{person}{Ravin Kumar}, \bibinfo{person}{Sarmad Hashmi}, \bibinfo{person}{Shubham Agrawal}, \bibinfo{person}{Zhitao Gong}, \bibinfo{person}{Jane Fine}, \bibinfo{person}{Tris Warkentin}, \bibinfo{person}{Ale~Jakse Hartman}, \bibinfo{person}{Bin Ni}, \bibinfo{person}{Kathy Korevec}, \bibinfo{person}{Kelly Schaefer}, {and} \bibinfo{person}{Scott Huffman}.} \bibinfo{year}{2024}\natexlab{}.
\newblock \bibinfo{title}{CodeGemma: Open Code Models Based on Gemma}.
\newblock
\newblock
\showeprint[arxiv]{2406.11409}~[cs.CL]
\urldef\tempurl%
\url{https://arxiv.org/abs/2406.11409}
\showURL{%
\tempurl}


\bibitem[Terragni et~al\mbox{.}(2020)]%
        {Terragni:gassert:ESECFSE:2020}
\bibfield{author}{\bibinfo{person}{Valerio Terragni}, \bibinfo{person}{Gunel Jahangirova}, \bibinfo{person}{Paolo Tonella}, {and} \bibinfo{person}{Mauro Pezz{\`e}}.} \bibinfo{year}{2020}\natexlab{}.
\newblock \showarticletitle{Evolutionary Improvement of Assertion Oracles}. In \bibinfo{booktitle}{\emph{Proceedings of the Joint Meeting on Foundations of Software Engineering}} \emph{(\bibinfo{series}{ESEC/FSE~'20})}. \bibinfo{publisher}{ACM}, \bibinfo{pages}{1178--1189}.
\newblock


\bibitem[Tillmann and Schulte(2005)]%
        {TillmanS2005}
\bibfield{author}{\bibinfo{person}{Nikolai Tillmann} {and} \bibinfo{person}{Wolfram Schulte}.} \bibinfo{year}{2005}\natexlab{}.
\newblock \showarticletitle{Parameterized unit tests}. In \bibinfo{booktitle}{\emph{ESEC/FSE 2005: Proceedings of the 10th European Software Engineering Conference and the 13th {ACM} {SIGSOFT} Symposium on the Foundations of Software Engineering}}. \bibinfo{address}{Lisbon, Portugal}, \bibinfo{pages}{253--262}.
\newblock


\bibitem[Tufano et~al\mbox{.}(2021)]%
        {Tufano:TestCase:2021}
\bibfield{author}{\bibinfo{person}{Michele Tufano}, \bibinfo{person}{Dawn Drain}, \bibinfo{person}{Alexey Svyatkovskiy}, \bibinfo{person}{Shao~Kun Deng}, {and} \bibinfo{person}{Neel Sundaresan}.} \bibinfo{year}{2021}\natexlab{}.
\newblock \bibinfo{title}{Unit Test Case Generation with Transformers and Focal Context}.
\newblock
\newblock
\showeprint[arxiv]{2009.05617}~[cs.SE]


\bibitem[Tufano et~al\mbox{.}(2022)]%
        {Tufano:Transformers:AST:2022}
\bibfield{author}{\bibinfo{person}{Michele Tufano}, \bibinfo{person}{Dawn Drain}, \bibinfo{person}{Alexey Svyatkovskiy}, {and} \bibinfo{person}{Neel Sundaresan}.} \bibinfo{year}{2022}\natexlab{}.
\newblock \showarticletitle{Generating Accurate Assert Statements for Unit Test Cases Using Pretrained Transformers}. In \bibinfo{booktitle}{\emph{Proceedings of the 3rd ACM/IEEE International Conference on Automation of Software Test}} (Pittsburgh, Pennsylvania) \emph{(\bibinfo{series}{AST '22})}. \bibinfo{publisher}{Association for Computing Machinery}, \bibinfo{address}{New York, NY, USA}, \bibinfo{pages}{54–64}.
\newblock
\showISBNx{9781450392860}
\urldef\tempurl%
\url{https://doi.org/10.1145/3524481.3527220}
\showDOI{\tempurl}


\bibitem[Wang et~al\mbox{.}(2018)]%
        {wang2018oracles}
\bibfield{author}{\bibinfo{person}{Chunhui Wang}, \bibinfo{person}{Fabrizio Pastore}, {and} \bibinfo{person}{Lionel Briand}.} \bibinfo{year}{2018}\natexlab{}.
\newblock \showarticletitle{Oracles for testing software timeliness with uncertainty}.
\newblock \bibinfo{journal}{\emph{ACM Transactions on Software Engineering and Methodology (TOSEM)}} \bibinfo{volume}{28}, \bibinfo{number}{1} (\bibinfo{year}{2018}), \bibinfo{pages}{1--30}.
\newblock


\bibitem[Wang et~al\mbox{.}(2004)]%
        {Wang:Oracles:Forte:2004}
\bibfield{author}{\bibinfo{person}{Xin Wang}, \bibinfo{person}{Ji Wang}, {and} \bibinfo{person}{Zhi-Chang Qi}.} \bibinfo{year}{2004}\natexlab{}.
\newblock \showarticletitle{Automatic Generation of Run-Time Test Oracles for Distributed Real-Time Systems}. In \bibinfo{booktitle}{\emph{Formal Techniques for Networked and Distributed Systems -- FORTE 2004}}, \bibfield{editor}{\bibinfo{person}{David de~Frutos-Escrig} {and} \bibinfo{person}{Manuel N{\'u}{\~{n}}ez}} (Eds.). \bibinfo{publisher}{Springer Berlin Heidelberg}, \bibinfo{address}{Berlin, Heidelberg}, \bibinfo{pages}{199--212}.
\newblock


\bibitem[Wang et~al\mbox{.}(2021)]%
        {Wang:few-shot-learning:csur:2022}
\bibfield{author}{\bibinfo{person}{Yaqing Wang}, \bibinfo{person}{Quanming Yao}, \bibinfo{person}{James~T. Kwok}, {and} \bibinfo{person}{Lionel~M. Ni}.} \bibinfo{year}{2021}\natexlab{}.
\newblock \showarticletitle{Generalizing from a Few Examples: {A} Survey on Few-shot Learning}.
\newblock \bibinfo{journal}{\emph{{ACM} Comput. Surv.}} \bibinfo{volume}{53}, \bibinfo{number}{3} (\bibinfo{year}{2021}), \bibinfo{pages}{63:1--63:34}.
\newblock
\urldef\tempurl%
\url{https://doi.org/10.1145/3386252}
\showDOI{\tempurl}


\bibitem[Watson et~al\mbox{.}(2020)]%
        {Watson:AssertionGenerationNN:ICSE:2020}
\bibfield{author}{\bibinfo{person}{Cody Watson}, \bibinfo{person}{Michele Tufano}, \bibinfo{person}{Kevin Moran}, \bibinfo{person}{Gabriele Bavota}, {and} \bibinfo{person}{Denys Poshyvanyk}.} \bibinfo{year}{2020}\natexlab{}.
\newblock \showarticletitle{On learning meaningful assert statements for unit test cases}. In \bibinfo{booktitle}{\emph{Proceedings of the ACM/IEEE 42nd International Conference on Software Engineering}}. \bibinfo{pages}{1398--1409}.
\newblock


\bibitem[Wei et~al\mbox{.}(2022)]%
        {Wei:Chain-of-Thought:2022}
\bibfield{author}{\bibinfo{person}{Jason Wei}, \bibinfo{person}{Xuezhi Wang}, \bibinfo{person}{Dale Schuurmans}, \bibinfo{person}{Maarten Bosma}, \bibinfo{person}{brian ichter}, \bibinfo{person}{Fei Xia}, \bibinfo{person}{Ed Chi}, \bibinfo{person}{Quoc~V Le}, {and} \bibinfo{person}{Denny Zhou}.} \bibinfo{year}{2022}\natexlab{}.
\newblock \showarticletitle{Chain-of-Thought Prompting Elicits Reasoning in Large Language Models}. In \bibinfo{booktitle}{\emph{Advances in Neural Information Processing Systems}}, \bibfield{editor}{\bibinfo{person}{S.~Koyejo}, \bibinfo{person}{S.~Mohamed}, \bibinfo{person}{A.~Agarwal}, \bibinfo{person}{D.~Belgrave}, \bibinfo{person}{K.~Cho}, {and} \bibinfo{person}{A.~Oh}} (Eds.), Vol.~\bibinfo{volume}{35}. \bibinfo{publisher}{Curran Associates, Inc.}, \bibinfo{pages}{24824--24837}.
\newblock
\urldef\tempurl%
\url{https://proceedings.neurips.cc/paper_files/paper/2022/file/9d5609613524ecf4f15af0f7b31abca4-Paper-Conference.pdf}
\showURL{%
\tempurl}


\end{thebibliography}

\end{document}